\documentclass[twocolumn,aps,prx,superscriptaddress]{revtex4-2}  

\usepackage{dcolumn}   
\usepackage{bm}        
\usepackage{amssymb}   
\usepackage{hyperref}
\usepackage{amsmath}
\usepackage{amssymb}
\usepackage{mathtools}
\usepackage{soul,xcolor}

\newcommand{\newparallel}{{\mkern3mu\vphantom{5}\vrule depth 0pt\mkern2mu\vrule depth 0pt\mkern3mu}}

\newcommand{\vect}[1]{\mathbf{#1}}
\newcommand{\citeasnoun}[1]{Ref.~\citenum{#1}}
\newcommand{\SupMat}{Supplemental Material \cite{supinfo}}

\newcommand{\figref}[1]{Fig.~\ref{#1}}
\newcommand{\figsref}[1]{Figs.~\ref{#1}}

 \newcommand{\secref}[1]{Sec.~\ref{#1}}

\DeclareMathOperator*{\argmin}{arg\,min}

\PassOptionsToPackage{normalem}{ulem}
\usepackage{ulem}

\usepackage{tikz,xcolor,hyperref}
\definecolor{lime}{HTML}{A6CE39}
\DeclareRobustCommand{\orcidicon}{
    \hspace{-2mm}
	\begin{tikzpicture}
	\draw[lime, fill=lime] (0,0) 
	circle [radius=0.16] 
	node[white] {{\fontfamily{qag}\selectfont \tiny ID}};
	\draw[white, fill=white] (-0.0625,0.095) 
	circle [radius=0.007];
	\end{tikzpicture}
	\hspace{-2.5mm}
}
\foreach \x in {A, ..., Z}{\expandafter\xdef\csname orcid\x\endcsname{\noexpand\href{https://orcid.org/\csname orcidauthor\x\endcsname}
			{\noexpand\orcidicon}}
}

\begin{document}

\title{Quasi-normal mode theory of the scattering matrix,\\ enforcing fundamental constraints for truncated expansions}

\author{Mohammed Benzaouia\orcidA} \email[Corresponding author: ]{medbenz@mit.edu} \affiliation{Department of Electrical Engineering and Computer Science, Massachusetts Institute of Technology, Cambridge, MA 02139, USA.}
\author{John D. Joannopoulos} \affiliation{Department of Physics, Massachusetts Institute of Technology, Cambridge, MA 02139, USA.}
\author{Steven G. Johnson\orcidB} \affiliation{Department of Mathematics, Massachusetts Institute of Technology, Cambridge, MA 02139, USA.}
\author{Aristeidis Karalis\orcidC} \email[Corresponding author: ]{aristos@mit.edu} \affiliation{Research Laboratory of Electronics, Massachusetts Institute of Technology, Cambridge, MA 02139, USA.}

\begin{abstract}
We develop a quasi-normal mode theory (QNMT) to calculate a system's scattering $S$~matrix, simultaneously satisfying both energy conservation and reciprocity even for a small truncated set of resonances. It is a practical reduced-order (few-parameter) model based on the resonant frequencies and constant mode-to-port coupling coefficients, easily computed from an eigensolver without the need for QNM normalization. Furthermore, we show how low-$Q$ modes can be separated into an effective slowly varying background response $C$, giving an additional approximate formula for $S$, which is useful to describe general Fano-resonant phenomena. We demonstrate our formulation for both normal and fixed-angle oblique plane-wave incidence on various electromagnetic metasurfaces.
\end{abstract}

\maketitle

\section{Introduction}

Scattering phenomena in all areas of wave physics are well described by the universal $S$-matrix operator. As the resonant (quasi-normal) modes (QNMs) of a system heavily determine its scattering response and coincide with the poles of $S$, numerous works~\cite{grigoriev2013optimization, krainov2016siegert, alpeggiani2017quasinormal, weiss2018qnmt, zhang2020quasinormal} have focused on expressing $S$ as an expansion over QNMs, calculated via eigensolvers. In this paper, we present a QNM theory (QNMT) for multiport lossless scatterers that \emph{simultaneously} satisfies all fundamental physical constraints of reciprocity, energy conservation and time-domain realness even for the practical case of a \emph{small truncated} QNM set (in contrast to previous formulations~\cite{alpeggiani2017quasinormal, weiss2018qnmt}) and without the need for the intricate normalization of the divergent QNMs~\cite{weiss2018qnmt, zhang2020quasinormal}. Weak absorption or gain can then be easily incorporated as a perturbation. Furthermore, by explicitly separating a slowly varying effective-background response $C$, we provide a novel additional formula for $S$, approximate but very convenient to design Fano-scattering systems~\cite{limonov2017fano} such as even-order elliptic filters~\cite{DESIGN}. This $C$ is calculated without resorting to any type of fitting~\cite{suh2004temporal, hsu2013bic, weiss2018qnmt} and without having to choose a specific background scattering medium~\cite{zhang2020quasinormal}: We simply use a subset of low-$Q$ modes of the entire system. We then build useful intuition for how various low-$Q$-mode configurations shape the background $C$. We demonstrate the accuracy of our QNMT for plane-wave incidence on several electromagnetic (microwave and photonic, 2-port, and 4-port) metasurfaces. In particular, we solve a nonlinear eigenproblem with a complex Bloch wave vector to calculate $S$ with QNMT for a fixed \emph{angle} of incidence (instead of a fixed transverse wave vector~\cite{vial2014quasimodal,Weiss2017angle}).

\medskip
The resonant modes of open physical systems are often called ``quasi-normal modes'' (QNM), as they are not square-integrable and exponentially diverge outside the resonator. The QNM eigenvalues correspond to the poles of physical quantities describing the system response. Based on a pole expansion of a desired such quantity, QNMT is usually concerned with identifying the pole residues/coefficients and any additional background terms. These expansions can allow a fast approximate solution for scattering and emission problems, while providing physical understanding and good spectral accuracy around sharp resonances, in contrast to direct numerical methods using frequency/time discretization~\cite{ching1998quasinormal, lalanne2018light, kristensen2020modeling, franke2020quantized, Settimi2003}. A quantity often studied is the Green's function~\cite{economou2006green}, which is an infinite-dimensional operator that can be used to construct any solution of the system. However, for problems with a finite number of scattering ports, the scattering $S$ matrix is usually the desired system descriptor, whose calculation is the goal of this paper. It is a simpler finite-dimensional operator which can be computed without requiring the full Green's function. For lossless 1-port systems, the numerator of the scalar $S$ is trivial, since its zeros simply coincide with the conjugates of the poles~\cite{nussenzveig1972causality}, while loss can be simply treated either by perturbation or by directly computing the zeros~\cite{grigoriev2013optimization, sweeney2020theory}. In the general multiport case, one recent approach used the full field equations to compute frequency-dependent $S$-expansion coefficients as volume integrals involving the QNMs and the excitation port fields to achieve good accuracy~\cite{zhang2020quasinormal}. Other QNMT formulations have shown how to project the Green's function onto the scattering ports to obtain the $S$ expansion, which is further given as a reduced-order model with frequency-independent residues, an advantage for simplicity and easier interpretation~\cite{krainov2016siegert, weiss2018qnmt}. Most of these derivations based only on field-equations solutions require the QNMs' normalization, which can be accomplished only by intricate techniques with increased computational complexity due to the QNM far-field divergence~\cite{kristensen2015normalization}. To avoid normalization, a new phenomenological approach was proposed in \citeasnoun{alpeggiani2017quasinormal}, starting from the coupled mode theory (CMT) equations~\cite{haus1984waves, suh2004temporal} and changing basis to the QNMs (a rather confusing approach, as the uncoupled orthonormalized modes required by CMT are ambiguous for arbitrary scatterers). However, for lossless reciprocal systems, these existing formulations do not guarantee energy conservation for a small truncated expansion, but presumably only in the infinite limit. While QNMTs with frequency-dependent expansion coefficients are expected to converge faster towards satisfying this important physical constraint, those with constant residues exhibit large errors for a practical small number of QNMs (\citeasnoun{weiss2018qnmt} mentions that they violate energy conservation visibly even with $301$ modes and we show examples where \citeasnoun{alpeggiani2017quasinormal} violates it by 50\% with few modes). In this paper, we consider an $S$-matrix expansion over the QNMs and directly derive conditions for it to satisfy the necessary physical constraints. We calculate the QNM-to-ports coupling matrix $D$ from simple surface integrals of the fields without need for QNM-normalization, and then finetune $D$ to prioritize and impose these conditions. In this way, rather than the expansion parameters being fixed from the field equations (e.g., if calculated via the Green's function), they are adjusted as more modes are included in order to enforce reciprocity and energy conservation for any finite sum. The final result is a simple equation for $S$ [Eq.~(\ref{phen_cmt})] using only the eigenfrequencies and the finetuned $D$ [Eq.~(\ref{optimD})] (\secref{sec-QNM-theory}). We confirm the improved accuracy of our QNMT using 2-port and 4-port electromagnetic metasurface examples, and with excitation of both normally and obliquely incident plane waves. For the latter, most previous approaches~\cite{vial2014quasimodal, Weiss2017angle} imposed a fixed incidence transverse wave vector ($k^\perp=\omega \textrm{sin}\theta/c$), so that the angle~$\theta$ changed with frequency~$\omega$ (given the constant wave speed $c$). Instead, QNMT can be used to compute $S(\omega)$ for fixed $\theta$ by evaluating the relevant QNMs involving eigenfrequency-dependent complex Bloch wave vectors, formulated as a generalized linear eigenproblem in \citeasnoun{Gras2019angle}, while we directly solve the nonlinear eigenproblem here (\secref{sec-examples}).

\medskip
In addition to providing a fast computational tool, QNMT (like CMT) has the advantage of offering a simple analytical model for gaining physical insight into resonant systems and for designing practical resonant devices. One interesting example is the case of Fano-resonant shapes~\cite{limonov2017fano} emerging from the interplay between a high-$Q$ resonance and a slowly varying background response, useful for sensors and filters. This background scattering is usually described by a separate matrix $C$, which previous works have almost always estimated only by fitting it \emph{a posteriori} to the total $S$, either with a polynomial approximation~\cite{weiss2018qnmt} or an effective averaged structure~\cite{suh2004temporal, hsu2013bic}. Recently, an exact volume-integral formula was alternatively derived by factoring out a choice of physical background~\cite{zhang2020quasinormal} (but may require further development to handle certain boundary conditions, such as perfect electric conductors in electromagnetism). While it is understood that this background is related to the low-$Q$ modes of the actual structure, a detailed systematic prescription to compute it directly from them and its relation to the final $S$ are lacking. In \secref{sec-background-C}, we extend our QNMT to non trivial direct-scattering pathways, by showing that a slowly varying $C$ can be calculated with our general recipe using only the system low-$Q$ modes and by placing the high-$Q$ modes into a different matrix $\bar{S}$, in order to then obtain a good approximation $S=\bar{S}C$ [Eq.~(\ref{S_TC})]. We then analyze simple low-$Q$ pole configurations corresponding to different physical interpretations of $C$, such as a desired background transmission or group delay. Finally, we demonstrate this additional formulation for the electromagnetic-metasurface examples mentioned above. 
\bigskip

\section{Quasi-Normal Mode Theory}
\label{sec-QNM-theory}

\subsection{Formulation}

We consider a general scattering problem of an arbitrary linear time-independent scatterer, coupled to incoming (excited) and outgoing (scattered) radiation via several physical linear ports. At the frequency $\omega$ of excitation, the scatterer has a total of $P$ ``coupling port modes'' (CPM) of radiation, which can be either single propagating modes of $P$ different physical ports or several propagating modes of fewer ports (while all other port modes are either evanescent, or of incompatible symmetry, or their coupling is simply too small at $\omega$). Let CPM $p$ propagate with wave vector $\mathbf{k}_p(\omega)$ and field $\bm{\phi}_p(\omega,\mathbf{r}) = \bm{\phi}_p^\perp(\omega,\mathbf{r}_p^\perp)e^{ik_p r_p^{\newparallel}}$, separable in the propagation ($r_p^{\newparallel}$) and transverse ($\mathbf{r}_p^\perp\perp\mathbf{k}_p$) coordinates. In the most common case of reciprocal lossless physical ports, the CPMs at $\omega$ are orthogonal under the standard (conjugated) ``power'' inner product (a cross-sectional overlap surface integral) and can be normalized to carry unit power $\left\langle\bm{\phi}_p^\perp|\bm{\phi}_q^\perp\right\rangle = \delta_{pq}$~\cite{haus1984waves}. Then, for the $P$ pairs of incident and scattered CPM waves, if the vectors $\mathbf{s}_{+}$ and $\mathbf{s}_{-}$ denote, respectively, their amplitudes at specific reference cross sections $r_p^{\newparallel}=z_p$ of their physical ports, $\left|s_{\pm p}\right|^{2}$ equals the power carried by the $\pm p$ wave, $\mathbf{s}_{\pm}^{\dagger}\mathbf{s}_{\pm}$ is the net incident/scattered power, and the system scattering matrix $S$ at these reference cross sections is defined by $\mathbf{s}_{-}=S \; \mathbf{s}_{+}$.

Since the scattering system is open (coupled to radiation), its Hamiltonian $H$ is non-Hermitian, so it supports a set of resonant modes [with resonant frequencies $\omega_n$ and fields $\bm{\psi}_n\left(\mathbf{r}\right)$ namely $H\left(i\omega_n\right)\bm{\psi}_n=i\omega_n\bm{\psi}_n$]. Causality and stability~\cite{OppenheimWillsky1983} (or simply passivity~\cite{nussenzveig1972causality}) imply that the system response is analytic in the upper half of the complex $\omega$ plane, namely $\omega_n$ must lie in the lower half plane. $\bm{\psi}_n$ are linearly independent but quasi-normal and non-orthogonal under the standard (conjugated) ``energy'' inner product (a volume integral). However, when the system is reciprocal, $H$ is complex symmetric, so its QNMs are orthogonal under the non-conjugated inner product $\left\{\bm{\psi}_n|\bm{\psi}_l\right\}=0$ for $n\neq l$ \citep{lalanne2018light, yan2018rigorous}. We consider $N$ such modes, whose normalization we leave \emph{unspecified}, and denote by the \emph{diagonal} matrix $\Omega$ their complex frequencies and by the vector $\mathbf{a}$ their amplitudes upon excitation. Moreover, it is usually assumed there are also pathways other than the resonant QNMs for direct scattering of input to output CPM waves, through the background medium without the scatterer, quantified by a separate scattering matrix $C$ \citep{haus1984waves, suh2004temporal}.

The part of the scattered field \emph{not} due to direct pathways ($\mathbf{s_{-}^{\textrm{Q}}} = \mathbf{s_{-}}-C\mathbf{s_{+}}$) can be written outside the scatterer as an expansion over the complete set of port modes (propagating CPMs and evanescent). QNMT makes the approximation that it can be written also within the volume $V$ of the scatterer as a linear combination of the $N$ QNMs (an assumption also used in CMT~\citep{suh2004temporal}): \bigskip{}
\begin{equation}\label{qnmt-approximation}
\mathbf{F}_\textrm{scat}^{\textrm{Q}}=\left\{
\begin{array}{l}
\sum\limits_{p=1}^{P}s_{-p}^{\textrm{Q}}(\omega)\bm{\phi}_p^\perp(\omega,\mathbf{r}_p^\perp)e^{ik_p(r_p^{\newparallel}-z_p)}\text{+evan.};\;\mathbf{r}\notin V\\
\sum\limits_{n=1}^{N}a_n(\omega)\bm{\psi}_n(\mathbf{r});\;\mathbf{r}\in V.
\end{array}\right.
\end{equation}

By inserting the second line into the exact equation for the field inside the scatterer and by mode matching the two lines on the cross-section $z'_p$ where the $p$ port meets the scatterer boundary, one respectively gets the final two QNMT equations, which, with $\mathrm{exp}\left(-i\omega t\right)$ notation, takes the form (for a rigorous derivation see, e.g.,, \citeasnoun{zhang2020quasinormal}):
\begin{equation}
\begin{array}{c}
-i\left(\omega-\Omega\right)\mathbf{a}=K^{t}\mathbf{s}_{+}\\
\mathbf{s}_{-}-C\mathbf{s}_{+}=D\mathbf{a},
\end{array}\label{general-cmt-eqns}
\end{equation}
where
\begin{equation}
D_{pn}(\omega)=e^{ik_p (z_p-z'_p)} \left\langle\bm{\phi}_p^\perp(\omega,\mathbf{r}_p^\perp)|\bm{\psi}_n(\mathbf{r}_p^\perp)\right\rangle_{z'_p}\label{D-definition}
\end{equation}
and $z_p-z'_p$ is the distance of the $p$-port reference cross-section from the boundary of the scatterer. This $z'_p$ cross-section choice for the calculation of the $D$ overlaps is further justified in \secref{C-from-port-shift}. The $P\times N$ matrices $K$ and $D$ quantify the couplings of the QNMs to the input and output CPM waves respectively, and they are generally frequency dependent. Equations.~(\ref{general-cmt-eqns}) with $\Omega$ diagonal (also known as state-space representation in diagonal canonical form in circuit theory~\cite{brogan1991modern}) constitute the basis of QNMT and their solution for the scattering matrix $S$ is given by
\begin{equation}
S=C-D(i\omega-i\Omega)^{-1}K^{t}.\label{general-cmt}
\end{equation}

Although a general $C\left(\omega\right)$ was included in Eqs.~(\ref{general-cmt-eqns}) and (\ref{general-cmt}) to align with literature, we rely on the presumed completeness of the resonant QNMs to stipulate that each incident CPM wave is scattered to other CPMs only due to resonances. Therefore, for now, we take $C$ as a diagonal phase matrix. Later, we will show that, indeed, low-$Q$ QNMs can be combined to write a general effective background $C\left(\omega\right)$; in particular, a fully-transmissive $C$ comes from a zero-frequency mode with an infinite radiative rate (\secref{sec_infinite-Gamma}).
\bigskip{}

\begin{figure}
\includegraphics[width=1\columnwidth,keepaspectratio]{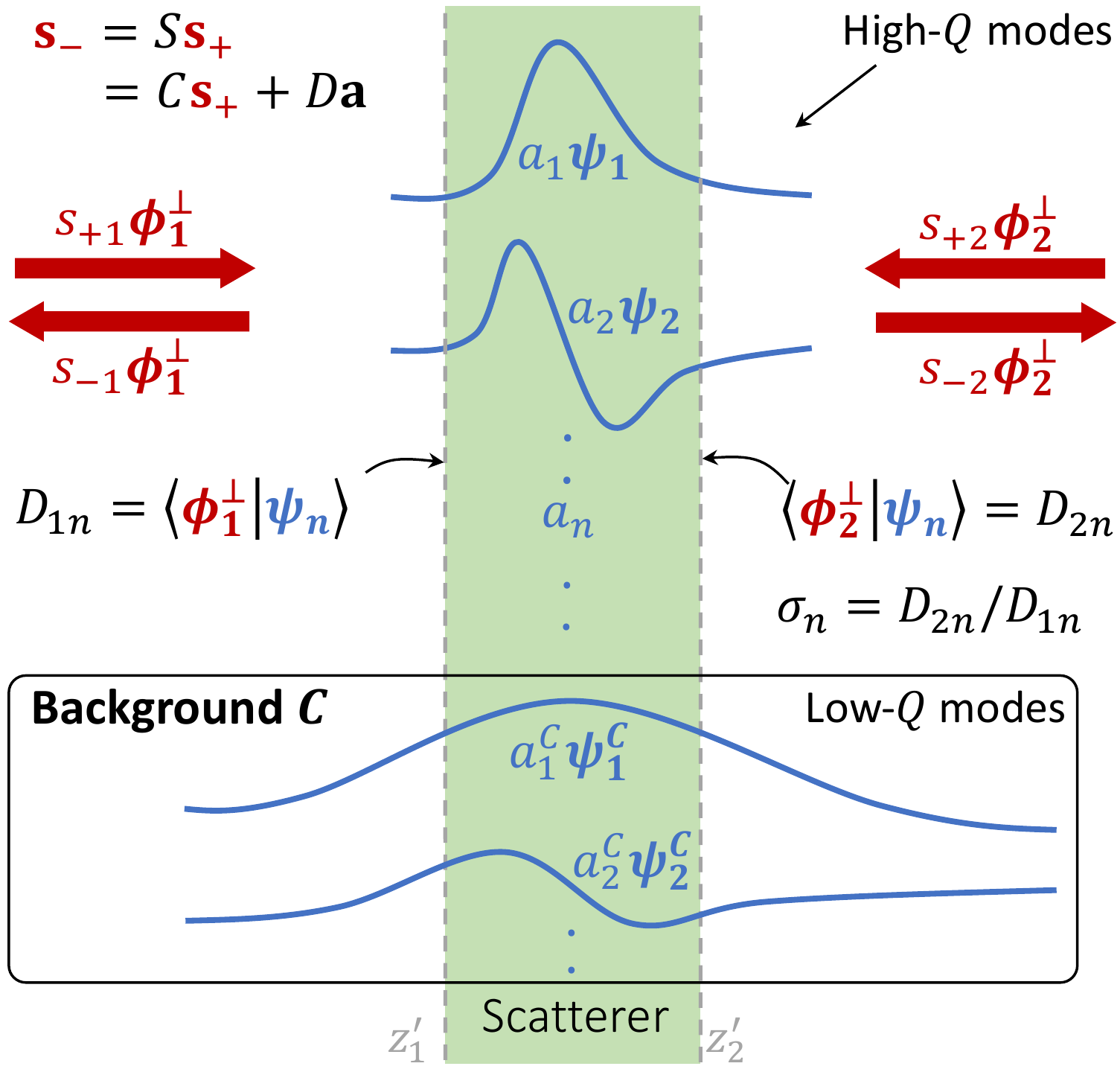} 
\caption{A 2-port scattering system with important QNMT parameters. At excitation frequency $\omega$, the coupling port modes (CPMs) with transverse fields $\phi_p^\perp$ have input and output amplitudes respectively $s_{\pm p}$, related by the $S$ matrix through $s_-=Ss_+$. The open scattering system supports quasi-normal modes (QNMs) with complex frequencies $\omega_n$ and fields $\psi_n$, which have amplitudes $a_n$ upon excitation. The CPM-to-QNM coupling coefficients are $D_{pn}$, with ratios $\sigma_n=D_{2n}/D_{1n}$. Imposing realness, unitarity, and symmetry constraints on $S$ allows us to compute it as a function of only $\omega_n$ and $\sigma_n$ [Eq.~(\ref{phen_cmt})]. Optionally, by separating low-$Q$ modes $\psi_n^C$, we can also construct a slowly-varying background matrix $C$, which can give a physical intuition about the scattering response and help in specific scattering designs~\cite{DESIGN}.}
\label{Fig-intro} 
\end{figure}

\paragraph*{Shifts of ports' reference cross sections---}

When $S(\omega)$ is a meromorphic function, it is useful to employ the Weierstrass factorization theorem and arguments from causality to write $S=e^{i\tau \omega}S'e^{i\tau \omega}$, where $\tau$ is a constant diagonal matrix with real positive elements and $S'$ is a ``proper'' rational function (Appendix~\ref{app_weierstrass}). Eq.~(\ref{general-cmt}) then implies that $C(\omega)=e^{i2\tau\omega}C'$, $D(\omega)=e^{i\tau\omega}D'$ and $K(\omega)=e^{i\tau\omega}K'$, where $C'$ is a diagonal constant phase matrix that can be taken equal to $-I$ (as we justify later and is often used in CMT~\cite{haus1984waves}) without loss of generality, and $D'$, $K'$ are now \emph{constant} matrices.

This is the case for CPMs with fields transverse to their direction of propagation ($\bm{\phi}_p\cdot\mathbf{k}_p=0$), such as plane waves or dual-conductor TEM microwave modes, which will be the focus of this paper. They have $k_p=\omega/c_p$ (where $c_p$ the wave velocity) and $\bm{\phi}_p^\perp(\mathbf{r}_p^\perp)$ independent of $\omega$, so Eq.~(\ref{D-definition}) suggests that $D_{pn}(\omega)=D'_{pn}e^{i\omega (z_p-z'_p)/c_p}$ with $D'_{pn}$ in fact constant. Thus, in all structures simulated in this paper, we remove this linear phase to compute $S'$, referenced at the new port cross sections $z'_p$ on the scatterer boundary. In practice, $\tau_{pp}$ may be slightly larger from $(z_p-z'_p)/c_p$, adding a small constant group delay just to the phases of $S$, so it is of no concern for applications dependent only on their amplitudes, such as amplitude filters~\cite{DESIGN}.

Therefore, hereafter we drop the $'$ and consider $S$ to be such a proper rational function that can be decomposed as
\begin{equation}
S=-[I+D(i\omega-i\Omega)^{-1}K^{t}]\Leftrightarrow S_{pq}=-\delta_{pq}-\sum_{n=1}^{N}\frac{D_{pn}K_{qn}}{i\omega-i\omega_n}.\label{S-expansion}
\end{equation}

For other types of CPMs, $S(\omega)$ may not be meromorphic, for example when higher-order CPMs have a cutoff frequency, which appears as a branch-point [and frequency dependent $D'(\omega)$, $K'(\omega)$]. Similarly to previous QNMT formulations with constant coefficients, such systems are not investigated in this work.

\bigskip{}

\paragraph*{Normalization independence---}

Recall that measurable physical quantities (such as the $S$ matrix) do not depend on the choice of normalization for the QNM amplitudes $\mathbf{a}$. For example, it is easily seen from the QNMT Eqs.~(\ref{general-cmt-eqns}) that, for the fixed normalization of $\mathbf{s}_{\pm}$, two different sets $\left(a_n,K_{qn},D_{pn}\right)$ and $\left(a'_n,K'_{qn},D'_{pn}\right)$ scale as $a'_n/a_n=K'_{qn}/K_{qn}=D_{pn}/D'_{pn}$, hence $S_{pq}$ in Eq.~(\ref{S-expansion}) is unchanged. Thus an overall scaling factor can be chosen arbitrarily for each QNM. 

For some physical quantities analytically computed via the QNM fields (such as the Green's function), this normalization independence is typically ensured by dividing with the volume-integral norm $\left\{\bm{\psi}_n|\bm{\psi}_n\right\}$ of these fields. However, since they are non-integrable, regularizing this norm is a procedure that adds complexity (e.g., choice of method and dependence on outgoing boundary condition)~\cite{kristensen2015normalization} to the QNMT formulations based only on calculations from the field equations~\cite{krainov2016siegert, weiss2018qnmt, zhang2020quasinormal}. In contrast, as highlighted in~\citeasnoun{alpeggiani2017quasinormal} and we show also later, this phenomenological (relying on physical constraints) QNMT formulation does not require such norm evaluation, so it is much simpler.

\subsection{Physical constraints}

We now proceed by imposing constraints on $S$, based on the physical properties of the system. \bigskip{}

\paragraph*{Realness---}
For many physical systems, real input fields lead to real output fields so that the system response $S(t)$ must also be real. In frequency domain, this realness of $S$ is stated as $S^{*}(i\omega)=S(-i\omega^{*})$. For such systems, the same relation holds for their Hamiltonian $H$ satisfying $H\left(i\omega\right)\bm{\psi}=i\omega\bm{\psi}$, so every QNM solution ($\omega_n$, $\bm{\psi}_n$) is paired with another QNM ($\omega_{n'}$, $\bm{\psi}_{n'}$) = ($-\omega_n^*$, $\bm{\psi}_n^*$). Similarly, the CPMs satisfy $\bm{\phi}^*_p(\omega) = \bm{\phi}_p(-\omega^*) \Rightarrow \bm{\phi}^{\perp*}_p = \bm{\phi}_p^\perp$. Equation~(\ref{D-definition}) thus shows that $D_{pn'}=D_{pn}^{*}$. Then, to satisfy $S^{*}(i\omega)=S(-i\omega^{*})$, Eq.~(\ref{S-expansion}) requires also $K_{pn'}=K_{pn}^{*}$. Note, $S$ realness implies that not only poles but also zeros of any $S_{pq}$ appear in pairs ($\omega_{o},-\omega_{o}^{*}$), and that $S_{pq}$ is a rational function of $i\omega$ with real coefficients. Note that our QNMT is still applicable to systems that do not satisfy realness (for example~\footnote{The resonances of the Schr{\"o}dinger equation do not appear in pairs in the $\omega=E/\hbar$ plane.}), where simply the QNMs included in the $S$-expansion do not appear in pairs. 
\bigskip{}

\paragraph*{Energy conservation---}

In absence of absorption or gain, energy conservation implies that the $S$ matrix is unitary ($\mathbf{s}_{+}^{\dagger}\mathbf{s}_{+} = \mathbf{s}_{-}^{\dagger}\mathbf{s}_{-}\Leftrightarrow S^{\dagger}S=I$)~\cite{haus1984waves}. In Appendix~\ref{app_unitarity}, we prove that a \emph{necessary and sufficient} condition is given by

\begin{equation}
K=D^{*}\left(M^t\right)^{-1}\textrm{, with }M_{nl}=\frac{\sum_{p=1}^{P}D_{pl}D_{pn}^{*}}{i\omega_{l}-i\omega_n^{*}}=M_{ln}^{*},\label{K-for-unitarity}
\end{equation}
thus $S$ can be written as

\begin{equation}
S=-\left[I+D(i\omega-i\Omega)^{-1}M^{-1}D^{\dagger}\right].\label{phen_cmt}
\end{equation}
In Appendix~\ref{app_real_fields}, we show that this Eq.~(\ref{K-for-unitarity}) choice of $K$ satisfies the realness requirement, namely, for $D_{pn'}=D_{pn}^{*}$, we get $K_{pn'}=K_{pn}^{*}$. In Appendix~\ref{app_uniqueness}, we rearrange Eq.~(\ref{phen_cmt}) to show that $S$ is fully and uniquely determined by the resonant frequencies $\omega_n$ and the ratios $\sigma_{r,pn}=D_{pn}/D_{r_{n}n}$ (for some chosen port $r_n$ for each mode $n$). These quantities can be readily calculated using any appropriate eigenmode solver, where $D_{pn}$ is determined by the surface integral in Eq.~(\ref{D-definition}) and the ratios $\sigma_{r,pn}$ remove the $\bm{\psi}_n$-normalization-dependent scaling-factor. Therefore, as promised, $S$ in Eq.~(\ref{phen_cmt}) does not require computing the volume-integral norms of the QNMs. Note also that, as more modes are included, the residue coefficients automatically update themselves through $M$ in order to satisfy energy conservation for the entire set, in contrast to other formulations based on the exact field equations, where these residues are cosntant~\cite{weiss2018qnmt, zhang2020quasinormal}. Finally, Eq.~(\ref{K-for-unitarity}) is different from the usual CMT expression of energy conservation $D^{\dagger}D=2\left|\mathrm{Im}\left\{ \Omega\right\} \right|$, associated with modes orthonormal under the standard ``energy'' inner product, which does not hold for QNMs.
\bigskip{}

\paragraph*{Reciprocity---}

Reciprocity implies that the $S$ matrix is symmetric ($ S=S^{t}$)~\cite{haus1984waves}. From Eq.~(\ref{S-expansion}), we can see that this is equivalent to having

\begin{equation}
\frac{K_{pn}}{D_{pn}}=\frac{K_{qn}}{D_{qn}}\Leftrightarrow K=D\Lambda\label{K-for-symmetry}
\end{equation}
and therefore

\begin{equation}
S=-\left[I+D(i\omega-i\Omega)^{-1}\Lambda D^{t}\right]\label{S-from-reciprocity}
\end{equation}
for some \emph{arbitrary} diagonal matrix $\Lambda$ with entries $\lambda_n$ [with the only restriction that $\lambda_{n'} = \lambda_n^{*}\neq0$, so that Eq.~(\ref{K-for-symmetry}) is compatible with realness], where a specific choice of $\lambda_n$ fixes the $\bm{\psi}_n$-normalization-dependent scaling-factor. For any such choice, as mentioned earlier, the numerator of pole $n$ for $S_{pq}$ in Eq.~(\ref{S-from-reciprocity}) stays the same, however, its calculation requires evaluation/regularization of divergent volume integrals involving the QNM fields (including their norm $\left\{\bm{\psi}_n|\bm{\psi}_n\right\}$), which we try to avoid here. Note also that, since $\Lambda$ can be arbitrary, the usually assumed condition of reciprocity $K=D$ (corresponding to the specific normalization choice $\Lambda=I$) is \emph{not necessarily} true.\bigskip{}

\paragraph*{$D$-optimization---}

In order to satisfy both energy conservation and reciprocity (unitarity
and symmetry of $S$), the input coupling coefficients $K$ must satisfy
both Eqs.~(\ref{K-for-unitarity}) and (\ref{K-for-symmetry}) simultaneously,
namely the output coupling coefficients $D$ must satisfy

\begin{equation}
D_{qn}\sum_{l}M_{\;nl}^{-1}D_{pl}^{*}=D_{pn}\sum_{l}M_{\;nl}^{-1}D_{ql}^{*}\Leftrightarrow D^{*}=D\Lambda M^{t}.\label{D-condition}
\end{equation}

Here again, for each resonance $n$, either $D_{r_{n}n}$ (for one port $r_n$) or $\lambda_n$ can be chosen arbitrarily. In practice, this reciprocity condition Eq.~(\ref{D-condition}) is a set of $PN$ equations. Let $D^c$ be the coupling coefficients computed from the eigenmode solver. In most cases, as we see in numerical examples, it turns out that $D^c$ are very close to satisfying this required condition, but they do not satisfy it perfectly, since the finite set of chosen resonances is not truly complete. This is why in \citeasnoun{alpeggiani2017quasinormal}, Eq.~(\ref{D-condition}) was not enforced exactly, rather the $N$ coefficients of $\Lambda$ were chosen as $\lambda_j = [X^\dagger X]_{jj}/[X^\dagger D]_{jj}$ with $X=D^* (M^t)^{-1}$ as one way to minimize its error (note~\footnote{$f(\Lambda)=|X\Lambda^{-1}-D|^2$ was minimized, however, different results would have been obtained for other choices, e.g., $f(\Lambda)=|X-D\Lambda|^2$ leads to $\lambda_j = [D^\dagger X]_{jj}/[D^\dagger D]_{jj}$.}), and their final $S$ matrix, which was formulated as in Eq.~(\ref{S-from-reciprocity}), was reciprocal but not necessarily unitary. Instead, here, we give priority to exactly satisfying these physical properties of the actual system, if we want our model to be a physically realistic and thus, as we will show, a more accurate description. Therefore, we  finetune the $PN$ coefficients $D$ by using a constrained optimization procedure, with the goal of exactly satisfying Eq.~(\ref{D-condition}) while staying as close as possible to the computed system performance:

\begin{equation}
D=\argmin_{D^{*}=D\Lambda M^{t}}f\left(D,D^c\right),\label{optimD}
\end{equation}
where $f(D,D^c)$ is some penalty function to ensure
that $D$ stays close to $D^c$ (note~\footnote{The optimization can for example be done using an augmented Lagrangian method \citep{conn1991globally}. Gradients can also be effectively computed using automatic differentiation \citep{RevelsLubinPapamarkou2016} or even analytically, when $\Lambda=I$.}). An obvious choice is $f(D,D^c) = \left\Vert D-D^c\right\Vert ^{2}$. Another option is $f(D,D^c) = \sum_{\{p,q\}}\left\Vert R_{pq}\left(D\right) - R_{pq}\left(D^c\right)\right\Vert^{2}$, where $R\left(D\right)$ represents the residues of the scattering matrix expansion given in Eq.~(\ref{phen_cmt}) and $\{p,q\}$ is a chosen subset of indices. When summing over $p\neq q$, the global optimum is reached for $D$ satisfying $R_{pq}\left(D\right) = R_{qp}\left(D\right) = \left[R_{pq}\left(D^c\right) + R_{qp}\left(D^c\right)\right]/2$, but such a set of $D$ is not guaranteed to exist. When it does exist, directly solving this system of equations has given the best results in our experience with actual 2-port systems.

\subsection{Properties of 2-port systems}

Since many common applications of scattering theory involve lossless reciprocal 2-port systems, we discuss some properties of their $S$ matrix.

Energy conservation leads to the unitary $S$ matrix of Eq.~(\ref{phen_cmt}), which is recast in normalization-independent form for a 2-port in Eq.~(\ref{eq:2-port-S21}). Realness requires that, for each mode ($\omega_n,\sigma_n$), where $\sigma_n=D_{2n}/D_{1n}$, the mode ($-\omega_n^{*},\sigma_n^{*}$) is also included. Then, the dependence of $S$ on $\sigma_n$ can be easily checked to satisfy, for $\gamma=\pm1$,
\begin{equation}
\begin{array}{c}
S_{11\{\omega_{n},\gamma\sigma_{n}\}}=S_{11\{\omega_{n},\sigma_{n}\}},\;
S_{12\{\omega_{n},\gamma\sigma_{n}\}}=\gamma^* S_{12\{\omega_{n},\sigma_{n}\}} \\
S_{21\{\omega_{n},\gamma\sigma_{n}\}}=\gamma S_{21\{\omega_{n},\sigma_{n}\}},\;
S_{22\{\omega_{n},\gamma\sigma_{n}\}}=S_{22\{\omega_{n},\sigma_{n}\}}.
\end{array}\label{eq:T-phase-dependence}
\end{equation}
[If only positive-frequency modes are considered and thus the realness requirement is relaxed---commonly known as a rotating-wave approximation (RWA)---Eq.~(\ref{eq:T-phase-dependence}) holds for any phase factor $\gamma=e^{i\varphi}$.] Moreover, swapping ports $1\leftrightarrow2$ corresponds simply to replacing $\sigma_n\leftrightarrow1/\sigma_n$:
\begin{equation}
S_{11\{\omega_{n},\sigma_{n}\}} = S_{22\{\omega_{n},1/\sigma_{n}\}}, \; S_{21\{\omega_{n},\sigma_{n}\}} = S_{12\{\omega_{n},1/\sigma_{n}\}}.
\label{eq:T-12-21}
\end{equation}
As a consequence, when all $\sigma_n$ are $\pm1$ (as for a symmetric structure, whose modes must be even or odd~\cite{JoannopoulosJo08-book}), $S_{11}=S_{22}$ and $S_{21}=S_{12}$, so the 2-port system is immediately reciprocal. [Obviously, $\sigma_n=\pm1$ is not \emph{necessary} for the reciprocity condition Eq.~(\ref{D-condition}) to hold.] Note that energy conservation alone implies that $\left|S_{12}\right|=\left|S_{21}\right|$, so reciprocity in lossless 2-ports is mostly a statement on the transmission phase responses.

Finally, in Appendix~\ref{app_zeros-2ports}, it is shown for a real lossless reciprocal 2-port system that (i) the zeros of $S_{21}$ can only appear as complex quadruplets  ($\omega_{o},\omega_{o}^{*},-\omega_{o},-\omega_{o}^{*}$), real or imaginary pairs ($\omega_{o},-\omega_{o}$), or at $\omega_{o}=0$ and (ii) for each zero pair ($\omega_{o},-\omega_{o}^{*}$) of $S_{11}$, ($-\omega_{o},\omega_{o}^{*}$) is a zero pair of $S_{22}$. The restrictions (i) on the $S_{21}$ zeros imply that its numerator is a polynomial of $\omega^{2}$ with real coefficients, optionally with multiplicative $i\omega$ factors. We emphasize that, for any lossless system with more than one port, the zeros of the $S$ coefficients are different from the ``$S$-matrix zeros'', where $\textrm{det}(S)=0$ and which always coincide with the conjugates of the poles~\cite{sweeney2020theory}.

\subsection{Absorption and gain}

In the presence of small absorption loss, $S$ is no longer unitary, but it can be calculated perturbatively when all relevant modes have high $Q$. In particular, the denominators of Eq.~(\ref{phen_cmt}) must obviously use the poles $\tilde{\Omega}$ of the actual absorptive system, however, if the QNM loss rates are split into radiative ($\Gamma_\mathrm{r}$) and non-radiative ($\Gamma_\mathrm{nr}$) parts, the numerators of Eq.~(\ref{phen_cmt}) scale as $DM^{-1}D \sim \Gamma_\mathrm{r}(1+\text{const}\cdot\Gamma_\mathrm{nr})$ [no scattering when $\Gamma_\mathrm{r}\rightarrow0$]. We then see that, for high-$Q$ modes (where both $\Gamma_{\mathrm{r}}$ and $\Gamma_{\mathrm{nr}}$ are much smaller than the frequencies), these numerators are equal to those of the lossless case to first order in $\Gamma_{\mathrm{r}},\Gamma_{\mathrm{nr}}$, namely absorption only affects the poles. (A similar argument is often implicitly used in CMT, where it is typically assumed that $\Gamma=D^{\dagger}D+\Gamma_{\mathrm{nr}}$ with $D$ not changing in the presence of $\Gamma_{\mathrm{nr}}$ because $\Gamma$ itself is small~\cite{alpeggiani2017quasinormal}, or is explicitly used to argue that the coupling coefficients to different CMT channels can be determined independently~\cite{JoannopoulosJo08-book}.) Therefore, in practice, to compute $S$ for absorptive or active scatterers, we first calculate the QNMs ($\omega_n,D_{pn}$) of the lossless (radiative only) structure and use them to evaluate the numerators of Eq.~(\ref{phen_cmt}). Then, we turn on absorption or gain mechanisms (adiabatically if needed for QNM-tracking purposes) to get the exact denominator poles ($\tilde{\omega}_n$). Since the lossless-case $D$ was finetuned for reciprocity, the non-unitary $S$ will still be symmetric. The perturbation argument assumes high-$Q$ modes but does not restrict the relative strength between radiation and absorption/gain rates. Indeed, as we see in the next examples, our QNMT gives quite accurate predictions even in the presence of modes with $\Gamma_\mathrm{nr}\gg\Gamma_\mathrm{r}$ and, in fact, even when relatively low-$Q$ modes are present.

\vspace{24pt}

\section{Examples in electromagnetism}
\label{sec-examples}

\vspace{16pt}

Our QNMT for the $S$ matrix is applicable to all kinds of wave physics, such as acoustics~\cite{Abom1991Smatrix,Sack2016acoustics,Tong2017acoustics}, electromagnetics~\cite{DESIGN, alpeggiani2017quasinormal, weiss2018qnmt, zhang2020quasinormal}, and quantum mechanics~\cite{datta1997electronic, krainov2016siegert}. Therefore, in our derivation, we used general physics-agnostic notation to render our results  usable for any wave-scattering problem. In this section, to examine the accuracy of our QNMT, we study multiple examples in electromagnetism.

\subsection{Normal incidence on microwave metasurface}

\begin{figure}
    \includegraphics[width=1\columnwidth]{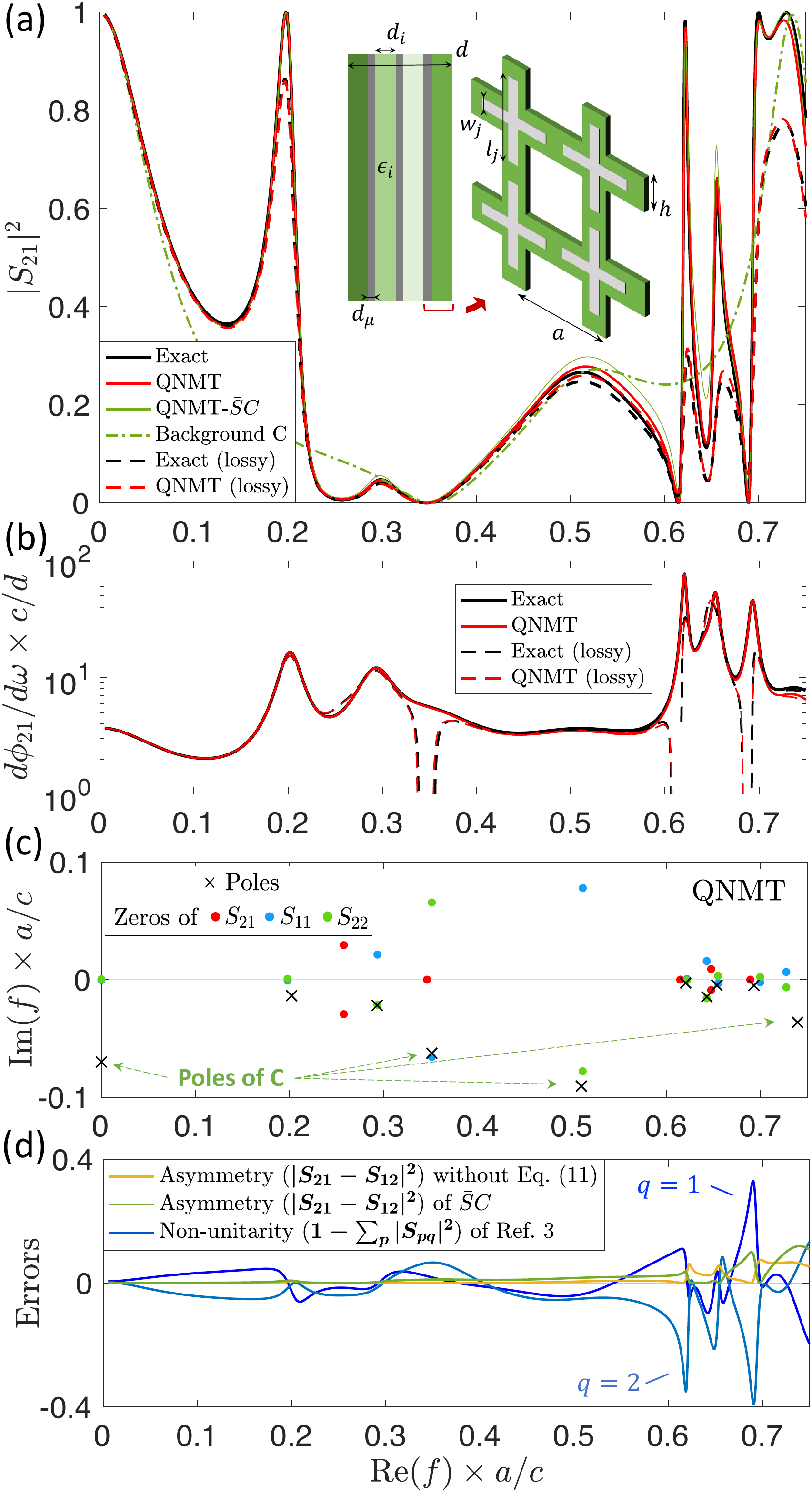} 
    \caption{QNMT modeling scattering of a plane wave normally incident from the left on the microwave metasurface depicted in the inset of (a). Parameters are: $a$ = 15mm, $h/a$ = 0.64, $w_j/a$ = (0.05, 0.2, 0.1), $l_j/a$ = (0.62, 0.92, 0.82), metal (grey) layers' thickness $d_{\mu}=18\textrm{\ensuremath{\mu}m}$, dielectric (green) layers' permittivities $\epsilon_i$ = (4, 6, 3, 10) and thicknesses $d_i/a$ = (0.1, 0.2, 0.3, 0.2). In the lossy simulation, we used copper and added loss $\tan\delta=0.01$ to all dielectric layers [for simplicity, $-\tan\delta$ is assumed for $\textrm{Re}(f)<0$ to maintain realness of $\epsilon(f)$]. Curves: (a) magnitude and (b) group delay of the transmission coefficient (where a constant delay of $0.24\times d/c$ was added to QNMT to match the exact simulation at low frequencies). (c) Lossless-system poles used in the QNM expansion (with their negative-frequency pairs used but not shown) and zeros of the resulting $S$ coefficients, confirming unitarity and symmetry of $S$. The dotted arrows point at the 4 modes used to compute the slowly varying background $C$ in the $S=\bar{S}C$ approximation. (d) Errors of asymmetric [Eq.~(\ref{phen_cmt}) without Eq.~(\ref{optimD}), and approximate Eq.~(\ref{S_TC})] or non-unitary [\citeasnoun{alpeggiani2017quasinormal}] QNMT formulations.}
    \label{Fig-test-structure}
\end{figure}

We now study scattering of an electromagnetic plane wave with frequency $f=\omega/2\pi$ normally incident on the metasurface depicted in the inset of \figref{Fig-test-structure}(a). It consists of alternating dielectric (green) and metallic (grey) layers, where the latter have been etched out to form two-dimensional square periodic lattices (of period $a$) of thin metallic crosses, whose centers are the same for all patterned layers. A square air-hole has also been etched throughout the entire thickness $d$ of the metasurface in the region between the crosses. The metal thickness is $18\,\textrm{\ensuremath{\mu}m}$ (corresponding to $0.5$ oz copper). We study for frequencies below the first diffraction cutoff ($f_\mathrm{cut}=c/a$ at normal incidence), so only transmission and reflection need to be considered. Moreover, there is $C_{2v}$ symmetry, so the response for normal incidence is independent of the polarization $\hat{e}$ and only 2 ports are needed. Numerical computation of the ``exact'' frequency response ($S$-matrix) for plane-wave excitation as well as of the eigenmodes for use in our QNMT is carried out using COMSOL Multiphysics~\citep{comsol}. Complex eigenfrequencies $\omega_n$ are immediately obtained from the eigensolver, while the coupling coefficients $D_{1n}$, $D_{2n}$ are computed from Eq.~(\ref{D-definition}) as ``power'' inner-product surface integrals
\begin{equation}
D_{pn}\propto\int_{z'_p}\left(\vect{E}_{p}^{*}\times\vect{H}_n+\vect{E}_n\times\vect{H}_{p}^{*}\right)\;\cdot d\mathbf{S}\label{D-overlap-integral}
\end{equation}
at the two (left/right for $p=1,2$) external boundaries of the metasurface between the QNM field $\bm{\psi}_n=\left(\vect{E}_n,\vect{H}_n\right)$ and the coupling port modes $\bm{\phi}_p^\perp=\left(\vect{E}_{p},\vect{H}_{p}\right)$ [plane waves in this case,  so $D_{pn} \propto \int_{z'_p} \left( \hat{e} \cdot \vect{E}_n \right) \; dS$],
where, as emphasized earlier, only their ratio $\sigma_n$ is needed. In Appendix~\ref{App:finite-element}, we provide further details and guidelines for the numerical simulations (mainly how to compute very-low-$Q$ modes), and within the \SupMat we give tables with the calculated QNMs ($\omega_n,\sigma_n$) for every structure presented in the paper.

For the asymmetric structure of \figref{Fig-test-structure}, the parameters were chosen arbitrarily to test a very general response, with $\sigma_n$ departing substantially from $\pm1$. Even so, we see a very good match between the exact numerical computation (black lines) and the QNM expansion of Eq.~(\ref{phen_cmt}) (red lines), for both the amplitude of $S_{21}=|S_{21}|e^{i\phi_{21}}$ (a) and its time delay $\tau_{21}=d\phi_{21}/d\omega \times c/d$ (b), in both cases of lossless (solid lines) and lossy (dashed lines) structures. In the lossless case, we emphasize again that, due to our symmetrization procedure of Eq.~(\ref{optimD}), the QNM expansion we obtained is both unitary and symmetric. This is why the zeros of $S_{21}=S_{12}$ are either real or complex conjugate pairs, and the zeros of $S_{11}$, $S_{22}$ complex conjugates of each other, as they should [\figref{Fig-test-structure}(c)]. The response with copper and dielectric losses mostly maintains the same overall features, and merely exhibits reduced transmission at high frequencies and ``superluminal'' ($0\leq\tau_{21}<1$) or negative group delay ($\tau_{21}<0$) around transmission zeros. The latter does not violate causality~\cite{Mojahedi2003timedelay}, instead it has been shown to necessarily occur at peaks of absorption~\cite{Bolda1993abnormalgv, daguanno2004density}, and thus is indeed typically associated with lossy bandstop (``notch'') transmission responses (such as zeros)~\cite{Ravelo2018ngd}. Our QNMT correctly predicts even these unusual phenomena.

To quantify the benefits of our QNMT, we calculate the errors associated with not exactly enforcing reciprocity or energy conservation, for the same QNMs of the lossless structure. If the $D$ coefficients are not finetuned with Eq.~(\ref{optimD}), $S$ from Eq.~(\ref{phen_cmt}) is not exactly symmetric, so \figref{Fig-test-structure}(d) shows the resulting error in $|S_{21}-S_{12}|^2$ (orange curve). It is relatively small (although increasing at higher frequencies), indicating that Eq.~(\ref{phen_cmt}) is already a good approximation. In contrast, the QNMT of \citeasnoun{alpeggiani2017quasinormal} [analogous to our Eq.~(\ref{S-from-reciprocity})] is reciprocal but violates energy conservation by large amounts, leading to non physical ``absorption/gain''. As indeed shown in \figref{Fig-test-structure}(d), for this lossless 2-port, the sum of transmission and reflection $|S_{pp}|^2+|S_{21}|^2$ ($p=1,2$) deviates from $1$ by almost $\pm 0.4$ at some frequencies (blue curves)! It turns out that these large errors exhibit themselves mostly in the reflection coefficients. Additionally, in both cases, the violation of a physical constraint leads to errors also in the group-delay prediction (such as negative group delay, which is impossible for a \emph{lossless} 2-port). However, these delay errors are of less importance, since they usually appear around transmission zeros and they are mitigated when a finite-bandwidth pulse is considered~\cite{peatross2000average}, as detailed in Appendix~\ref{App:group-delay}.

One key advantage of the QNMT method is that it resolves spectra around very-high-$Q$ modes with perfect detail, while a frequency simulation requires a very dense uniform frequency grid to resolve them, being ignorant of their location. This, in turn, leads to a stark benefit in speed for QNMT. For this example, on the same machine and finite-element mesh, the QNMT calculation took an average of $\sim 60$ secs per mode ($\times 10$ modes in Fig.~\ref{Fig-test-structure}), while the frequency-domain calculation an average of $\sim 100$ secs per point ($\times 600$ points in Fig.~\ref{Fig-test-structure}).

\vspace{12pt}

\subsection{4-port metasurface via coupled polarizations}

\vspace{12pt}

\begin{figure}[!h]
\includegraphics[width=1\columnwidth]{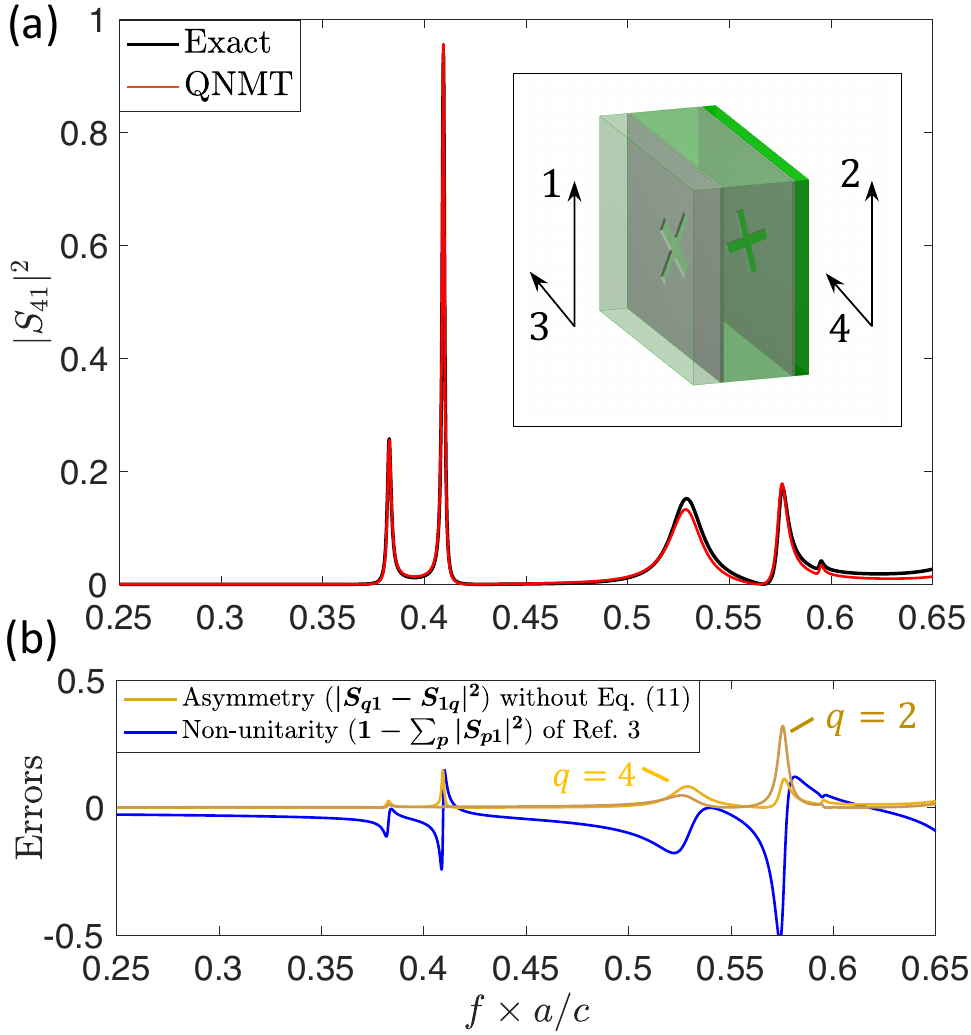} \caption{QNMT modeling of cross-polarization transmission of a normally incident plane wave from a metasurface with rotated apertures. The two planar metallic sheets are periodically patterned (period $a=15mm$) with same cross-like apertures [with widths $(0.05a,0.1a)$ and lengths $(0.5a,0.3a)$ in two orthogonal directions], which are rotated by angles $30^\circ$ and $60^\circ$ with respect to the polarization of port 1. The three uniform dielectric layers have $\epsilon_i = (2,3,5)$ and thicknesses $d_i/a=(0.1,0.05,0.15)$. (b) Errors of asymmetric [Eq.~(\ref{phen_cmt}) without Eq.~(\ref{optimD})] or non-unitary [\citeasnoun{alpeggiani2017quasinormal}] QNMT formulations.}
\label{Fig-4-port}
\end{figure}

We now consider the 4-port system described in \figref{Fig-4-port} which consists of a microwave metasurface with three dielectric layers sandwiching two metallic sheets, with patterned arrays of \emph{rotated} cross-like apertures. The ports correspond to the two polarizations on the left (1,2) and right (3,4) sides of the structure. This system does not have the required symmetry for the normally incident plane-wave polarization to be conserved, instead the two orthogonal polarizations on each side cross-couple in both reflection and transmission, so a 4-port system is needed.

In Fig.~\ref{Fig-4-port}(a), we plot the cross-polarization transmission $\left|S_{41}\right|^2$ and again find a good agreement with our QNMT. In the \SupMat, we show the other components of the $S$-matrix. In Fig.~\ref{Fig-4-port}(b), we again show for comparison the errors of QNMT without symmetry or unitarity. For the non-unitary QNMT of \citeasnoun{alpeggiani2017quasinormal}, the quantity $1-\sum_p|S_{p1}|^2$ reaches values as low as $-0.5$ (blue curve), in stark contradiction with energy conservation. The asymmetric QNMT using Eq.~(\ref{phen_cmt}) without the $D$ correction of Eq.~(\ref{optimD}) has errors in $|S_{pq}-S_{qp}|^2$ as large as 0.3 (orange curves). Moreover, in this example, it also has nonzero $|S_{pq}|^2-|S_{qp}|^2$ with an error up to $\sim$ 0.015. This happens, because, for $P$-port systems with $P>2$, unitarity alone does not guarantee $\left|S_{pq}\right|=\left|S_{qp}\right|$ anymore, so our method of fine-tuning $D$ to also enforce symmetry [Eq.~(\ref{optimD})] corrects errors not just in the scattering phase, but in the amplitudes too.

\newpage
\subsection{Oblique incidence on 2d photonic metasurface}

When a plane wave is incident on a metasurface at an angle $\theta$, its transverse wave vector component at frequency $\omega$ is $k^\perp = \omega\; \mathrm{sin}\theta/c$. Phase matching then imposes that this must also be the Bloch wave vector within the metasurface. QNMT modeling with such excitation may seem intractable at first glance, if one tries to obtain a full band diagram to apply QNMT at each fixed real $k^\perp$. However, we calculate here the relevant QNMs from a nonlinear eigenproblem, where the phase matching condition is imposed at the complex eigenfrequency by analytic continuation (thus giving a complex Bloch wave vector $k_n^\perp=\omega_n\; \mathrm{sin}\theta/c$). To find such unusual resonances, we developed software for two-dimensional (2d) dielectric structures, whose geometry can be split into uniform layered sections: at any complex $\omega$, these complex Bloch modes are calculated within each section with a $T$-matrix formulation, then matched at interfaces between sections, and finally radiation conditions are applied to find the resonances (similar to CAMFR~\cite{CAMFR} and other interface mode-matching analyses~\cite{dossou2012modematch,paul2011modematch}).

\begin{figure}[!htp]
\includegraphics[width=1\columnwidth]{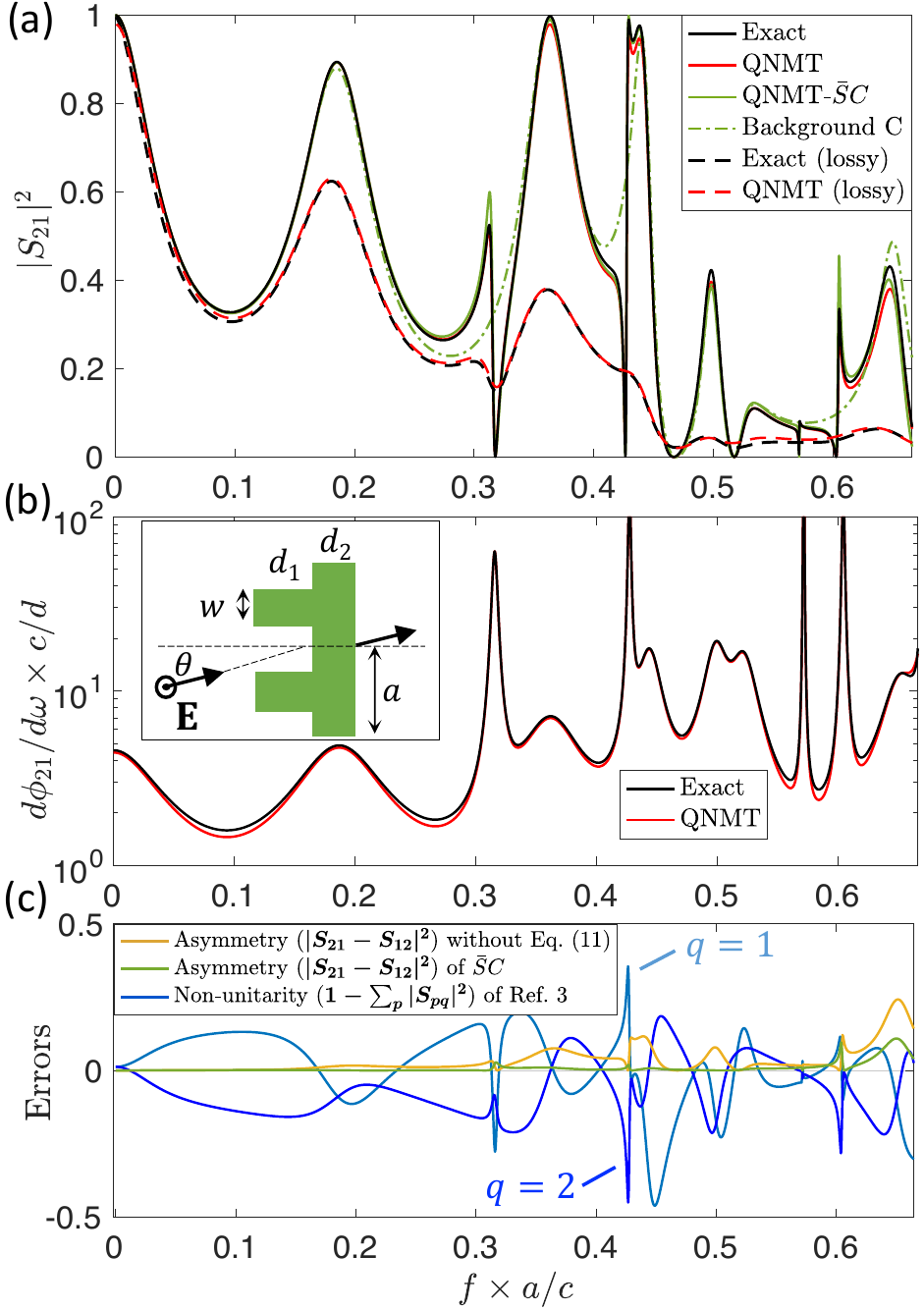} \caption{QNMT modeling of transmission of obliquely incident ($\theta=30^{\circ}$) TE plane wave from a dielectric ($\epsilon=11$) grating with $w/a=0.4$ and $d_i/a=(0.6,0.4)$ so $d=d_1+d_2=a$. The lossy structure has $\epsilon=11+0.77i$ (loss tangent $\tan\delta=0.07$).}
\label{Fig-angle}
\end{figure}

We then study scattering of a plane wave incident at a $30^{\circ}$ angle on a 2d photonic grating with its $\mathbf{E}$-field transverse to the plane [inset of Fig.~\ref{Fig-angle}(a)]. For frequencies below the first diffraction cutoff $f_\mathrm{cut}=c/a(1+\mathrm{sin}\theta)$, the system again has only 2 ports. In Fig.~\ref{Fig-angle}(a,b), we show transmission, calculated both exactly (black curves) and with our QNMT (red curves). The agreement is indeed very good all throughout the range. The QNMTs without unitarity or symmetry reach errors $\sim 0.5$ and $\sim 0.25$ respectively [Fig.~\ref{Fig-angle}(c)]. We highlight that QNMTs are expected to improve as more (higher-frequency) modes are included, however, the non-unitary formulation~\cite{alpeggiani2017quasinormal} exhibits large errors even down to very low frequencies. Transmission is also plotted in the case of strong dielectric losses and it highlights that our perturbative approach works very well, even though now (due to absorption) several modes have decay rates $\tilde{\Gamma}_n$ more than an order of magnitude larger than their rates $\Gamma_n$ for the lossless structure (see QNMs within \SupMat).

\section{Background scattering representation by low-$Q$ modes}
\label{sec-background-C}

In some design situations~\cite{DESIGN}, one needs an effective slowly-varying background response, which has to be designed collectively and in conjunction with the high-$Q$ modes. This background response (responsible also for the concept of Fano resonances~\cite{limonov2017fano}) is usually modeled in standard CMT via the direct-coupling matrix $C$ in Eq.~(\ref{general-cmt-eqns}). In most cases, researchers have approximated $C$ by simulating an effective background structure derived by some type of topology averaging which removes the high-$Q$ resonances, with parameters chosen \emph{a posteriori} for a best fit \citep{suh2004temporal, hsu2013bic, weiss2018qnmt}. Here, we show how $C$ can be calculated using the actual structure under study by appropriately combining its low-$Q$ modes, providing also intuition for this dependency and its physical interpretation. Given the background $C$, we also derive how the total $S$ can be computed and we test it for the electromagnetic examples of  \secref{sec-examples}.

\subsection{QNMT with background $C$ matrix}

Consider a lossless system supporting some high-$Q$ resonant modes ($\omega_n=\Omega_n-i\Gamma_n,\;D_n$), while the rest ($\omega_n^{C}=\Omega_n^{C}-i\Gamma_n^{C},\;D_n^C $) have much smaller $Q$ (\figref{Fig-intro}). Starting from Eq.~(\ref{S-expansion}), we combine these low-$Q$-mode terms within the sum to define $C\left(\omega\right) \equiv S_{\{\omega_n^C,D_n^C\}}\left(\omega\right)$, so that $S$ takes the form of Eq.~(\ref{general-cmt}), with $\Omega$ including only the high-$Q$ modes. In the limit $\Gamma^{C}\rightarrow \infty$, $C$ becomes a frequency-independent matrix, which is unitary, since $S(\omega\rightarrow\infty) = C$. Therefore, we can use Eqs.~(\ref{phen_cmt}) to calculate $C$ and Eq.~(\ref{optimD}) to guarantee its symmetry. In this process, $M_{nl}^{C} \approx \sum_{p=1}^{P}D_{pl}^{C}D_{qn}^{C*}/\left(\Gamma_{l}^{C}+\Gamma_n^{C}\right)$, so the $\Gamma_{n}^{C}\rightarrow\infty$ cancel out between $M^{C}$ and the $C$-denominator poles, leading to a constant $C$ not necessarily equal to $-I$. In the next subsection, we calculate this limiting $C$ for some simple but useful pole configurations. Now, for the total $S$-matrix of Eq.~(\ref{general-cmt}), following the same procedure of Appendix~\ref{app_unitarity} but with $-I$ replaced by a unitary symmetric constant $C$, one can easily see that $S^{\dagger}S=I$ is now equivalent to $K^{t}=-M^{-1}D^{\dagger}C$. Therefore, the background $C$ can be factored out of Eq.~(\ref{general-cmt}) to write:

\begin{equation}
\begin{split}
S=\bar{S}&C =\left[I+D(i\omega-i\Omega)^{-1}M^{-1}D^{\dagger}\right]C\\
\Leftrightarrow S_{pq}=&\; C_{pq}+\sum_{n}\frac{D_{pn}\sum_{l}M_{\;nl}^{-1}\sum_{r}D_{rl}^{*}C_{rq}}{i\omega-i\omega_n}.
\end{split}\label{S_TC}
\end{equation}
Here, $\bar{S}$ has the form of a separate scattering matrix, which itself also satisfies realness, unitarity and the properties of Eqs.~(\ref{eq:T-phase-dependence}, \ref{eq:T-12-21}). The condition of Eq.~(\ref{D-condition}), to additionally impose exact symmetry on $S$, is modified to
\begin{equation}
\begin{split}
D_{qn}\sum_{l}M_{\;nl}^{-1}\sum_{r}D_{rl}^{*}C_{rp}&=D_{pn}\sum_{l}M_{\;nl}^{-1}\sum_{r}D_{rl}^{*}C_{rq}\\
\Leftrightarrow CD^*=&-D\Lambda M^t,
\end{split}\label{D-condition-withC}
\end{equation}
through which a finetuned $D$ can be evaluated by an optimization procedure, as before, to obtain the final $S$. 

As we discuss in Appendix~\ref{App:finite-element}, such extremely-low-$Q$ modes are very difficult to locate numerically. Thankfully, in practice, real systems usually have modes with reasonably low  $Q$, which are easier to find. However, the problem then is that their associated $C\left(\omega\right)$ is slowly varying instead of constant and is \emph{not} necessarily unitary (as it does not describe a physical system by itself). Nevertheless, we can still use Eq.~(\ref{phen_cmt}) for those low-$Q$ modes to obtain a unitary (thus approximate) $C$, symmetrize it with Eq.~(\ref{optimD}), and then use Eq.~(\ref{S_TC}) with the high-$Q$ modes inside $\bar{S}$ to get $S$ simply as a best-effort \emph{approximation} to the actual system response. This $\bar{S}C$ construction still guarantees that $S$ will also satisfy realness [$S^{*}(\omega)=\bar{S}^{*}(\omega)C^{*}(\omega)=\bar{S}(-\omega^{*})C(-\omega^{*})=S(-\omega^{*})$] and unitarity ($S^{\dagger}S=C^{\dagger}\bar{S}^{\dagger}\bar{S}C=I$), but does \emph{not} guarantee reciprocity ($S^{t}=C^{t}\bar{S}^{t}\neq \bar{S}C=S$) even though $C^t=C$ [at least, unitarity implies $\left|S_{12}\right|=\left|S_{21}\right|$ for a 2-port]. Attempts to symmetrize $S$ (for example, \footnote{One could, for example, use the approximation that, within the bandwidth of each high-$Q$ resonance in Eq.~(\ref{S_TC}), $C(\omega)\approx C(\omega_n)$, which translates to substituting $C_{pq}\rightarrow C_{pq,n}$ in the symmetrization condition Eq.~(\ref{D-condition-withC}). However, this only gives an approximate condition that does not improve much on the original result.}) are not expected to have much success: if it were possible to exactly satisfy also reciprocity for a varying $C(\omega)$, one would then be able to build a unitary and symmetric $S$ by multiplying unitary and symmetric $C$ matrices formed by individual modes, which obviously is not possible. Regardless, as we show later in examples, for many physical systems, $S=\bar{S}C$ is a good enough approximation, which we use in separate work~\cite{DESIGN} to design accurate metasurface standard (e.g., elliptic) filters with a non-trivial background. In such design situations where a specific background is desired, $C=\bar{S}^{-1}S$ can alternatively be used to estimate and then design $C(\omega_c)$ at the target frequencies $\omega_c$ without having to calculate any low-$Q$ modes, rather by using $S(\omega_c)$ from a direct simulation and $\bar{S}(\omega_c)$ from QNMT using only the high-$Q$ modes.

\subsection{$C$ matrix due to $\Gamma\rightarrow\infty$ modes in 2-port systems}
\label{sec_infinite-Gamma}

In this subsection, we study some very basic configurations of high-$\Gamma$ modes $(\omega_n^C,\sigma_n^C)$ in 2-port systems to build intuition on their influence in shaping the background scattering. (For notational simplicity, here we drop the superscript ``$^C$''.)
\bigskip{}

\paragraph*{Single zero-frequency mode $(-i\Gamma_\mathrm{o},\sigma_\mathrm{o})$---}
Consider a 2-port system supporting a zero-frequency mode $\omega_\mathrm{o}=-i\Gamma_\mathrm{o}$. As explained above, when $\Gamma_\mathrm{o}\rightarrow\infty$, this mode can be factored out of $S$ into a unitary symmetric background response $C$. The symmetry of $C$ dictates that the modal ports-coupling ratio $\sigma_\mathrm{o}$ is \emph{real}. Using Eq.~(\ref{eq:2-port-S21}) for just this mode, we find $M=(1+\sigma_\mathrm{o}^2)/2\Gamma_\mathrm{o}$ and then
\begin{equation}
\begin{split}
C_{pq} =-\delta_{pq}&-\frac{\sigma_{p\mathrm{o}}M^{-1}\sigma_{q\mathrm{o}}}{i\omega-\Gamma_\mathrm{o}}\xrightarrow{\Gamma_\mathrm{o}\gg|\omega|}-\delta_{pq}+\frac{2\sigma_{p\mathrm{o}}\sigma_{q\mathrm{o}}}{1+\sigma_\mathrm{o}^2} \\
\Leftrightarrow\: C = &\begin{pmatrix}r & t\\ t & -r \end{pmatrix} ; \;r=\frac{1-\sigma_\mathrm{o}^2}{1+\sigma_\mathrm{o}^2},\;t=\frac{2\sigma_\mathrm{o}}{1+\sigma_\mathrm{o}^2},
\end{split}\label{eq:C-single-mode}
\end{equation}
which can give any ``reflection matrix'' in the orthogonal group $O(2)$. A given transmission $t$ is achieved for
\begin{equation}
\sigma_\mathrm{o}=\frac{1}{t}\pm\sqrt{\frac{1}{t^{2}}-1}.
\label{eq:s0-from-tr}
\end{equation}
In particular, we obtain $r=1$ for $\sigma_\mathrm{o}=0$, while $r=-1$ for $\sigma_\mathrm{o}\rightarrow\infty$. On the other hand, $\sigma_\mathrm{o}=\pm1$ gives a fully-transmissive background with $t=\pm1$.

A concrete example which confirms this last result comes from considering a uniform material slab with thickness $d$ and refractive index $\tilde{n}$. Its ``Fabry--Perot'' modes are an equispaced spectrum given by~\cite{lalanne2018light}
\begin{equation}
\omega_{n}=\frac{2c}{\tilde{n}d}\left(\frac{n\pi}{2}-i\cdot\textrm{atanh}\frac{1}{\tilde{n}}\right),\;\sigma_n=\left(-1\right)^n.\label{eq:Fabry-Perot-modes}
\end{equation}
Consider now appropriately large $\tilde{n}$, so that $\textrm{atanh}\left(1/\tilde{n}\right) \ll \pi/2\Leftrightarrow\Gamma_\mathrm{o}\ll\Omega_1$. Then, at frequencies of interest $0<\omega\ll\Omega_1$, the modal contribution ($\sim\Gamma_\mathrm{o}/\Omega_n$) is negligible for all $n\neq0$. In the limit $d\rightarrow0$, indicating the absence of slab, the only relevant $n=0$ system mode has $\Gamma_\mathrm{o}\rightarrow\infty$. Therefore, perhaps counter-intuitively, full transmission can be seen as equivalent to such a mode at $0-i\infty$ with $\sigma_0=1$.

This result of $t=1$ for a zero-thickness slab is a consequence our initial phase choice $C'=-I$ in Eq.~(\ref{S-expansion}). If we had instead chosen $C'=+I$, we would have obtained $t=-1$ for zero thickness, which would be a valid but awkward phase convention.
\bigskip{}

\paragraph*{Conjugate-modes pair $[(\omega_\mathrm{o},\sigma_\mathrm{o}), (-\omega_\mathrm{o}^*,\sigma_\mathrm{o}^*)]$---}
Let us now consider a single mode $\omega_\mathrm{o} = \Omega_\mathrm{o}-i\Gamma_\mathrm{o}$ with $\Omega_\mathrm{o} \neq 0$, together with its negative (paired) mode at $-\omega_\mathrm{o}^*$. The symmetry of the associated background matrix $C$ again requires a real $\sigma_\mathrm{o}$, and Eq.~(\ref{eq:2-port-S21}) gives
\begin{equation}
\begin{split}
C_{pq} =& \; -\delta_{pq}-\frac{2\sigma_{p\mathrm{o}}\sigma_{q\mathrm{o}}}{1+\sigma_\mathrm{o}^2}\frac{i2\Gamma_\mathrm{o}\omega}{[i(\omega-\Omega_\mathrm{o})-\Gamma_\mathrm{o}][i(\omega+\Omega_\mathrm{o})-\Gamma_\mathrm{o}]}\\
&\Rightarrow \; C_{pq}(\Omega_\mathrm{o}) = -\delta_{pq} +\frac{2\sigma_{p\mathrm{o}}\sigma_{q\mathrm{o}}}{1+\sigma_\mathrm{o}^2}\frac{1}{1+i\Gamma_\mathrm{o}/2\Omega_\mathrm{o}}.
\end{split} \label{eq:C-conjugate-pair}
\end{equation}
When $|\omega-\Omega_\mathrm{o}|\ll\Gamma_\mathrm{o}\ll2\Omega_\mathrm{o}$ (so that RWA holds), $C(\omega\sim\Omega_\mathrm{o})$ again takes the single-mode value of Eq.~(\ref{eq:C-single-mode}) independent of $\Omega_\mathrm{o}$. Instead, when $\Gamma_\mathrm{o}\gg2\Omega_\mathrm{o}$, the two broad resonances effectively cancel each other and $C\approx-I$.
\bigskip{}

\paragraph*{Two uncoupled modes $[(\omega_1,\sigma_1),(\omega_2,\sigma_2)]$---}
Two modes with $\Omega_{1,2}\neq0$ and $\sigma_1\sigma_2^*=-1$ do not couple [$M_{12}=0$ in Eq.~(\ref{eq:2-port-S21})] and their pole contributions add up independently in $C(\omega)$. Symmetry of their $C$ again forces $\sigma_1=-1/\sigma_2$ to be real. [This is the case of an even ($\sigma_1=1$) and an odd ($\sigma_2=-1$) mode. Another example with $\sigma_1\rightarrow 0^\pm$ and $\sigma_2\rightarrow \mp\infty$ can occur for a strongly reflecting mirror, where each mode is localized on one of the two asymmetric sides and couples mainly to one port, thus defining two essentially disjoint 1-port systems.] Under the RWA $|\omega-\Omega_{1,2}| \ll \Gamma_{1,2}\ll2\Omega_{1,2}$, the pole contribution of each mode is like the one in Eq.~(\ref{eq:C-single-mode}) and then $C(\omega\sim\Omega_{1,2})\approx I$. However, in the limit $\Gamma_{1,2}\gg2\Omega_{1,2}$, their negative poles cannot be ignored, the contributions are as in Eq.~(\ref{eq:C-conjugate-pair}), and the result is instead $C\approx-I$.
\bigskip{}

\paragraph*{Equispaced (Fabry-Perot) modes $[n\omega_\mathrm{o}-i\Gamma_\mathrm{o},(-1)^n]$---}
When the structure has localized resonant elements but does not exhibit very strong overall reflection, the background response is commonly assumed to arise from the averaged geometry. When this is a simple uniform material slab, it corresponds to a Fabry-Perot system with the infinite set of equispaced alternating-symmetry modes given in Eq.~(\ref{eq:Fabry-Perot-modes}) and a ``comb'' transmission response (see e.g., \figref{Fig-Dboundary} blue circles and dashed lines). In the limit $\tilde{n}\rightarrow1$, the system approaches a slab of free space of thickness $d$ and its array of modes (equi-spaced by $\pi c/d$) is shifted down towards $-i\infty$. The corresponding limiting value of $C$ is simply the scattering matrix for propagation through $d$, namely $|C_{21}|\rightarrow1$ and $d\phi_{21}^C/d\omega\rightarrow d/c$.
\medskip{}

\paragraph*{One-sided equispaced modes $[n\omega_\mathrm{o}-i\Gamma_\mathrm{o},0\emph{ or }\infty]$---}
When instead the structure has strongly reflective components separating the two ports, the background response will comprise low-$Q$ modes with fields mostly localized on either of the two port sides, thus having $|\sigma_n|\ll1$ or $|\sigma_n|\gg1$ respectively, adding up to $|C_{11}|,|C_{22}|\approx1$. If the average geometry on one side is a uniform material slab of thickness $d/2$ on a perfect mirror, its modes will again be Eq.~(\ref{eq:Fabry-Perot-modes}) but for only odd or only even $n$. As $\tilde{n}\rightarrow1$ and $\Gamma_\mathrm{o}\rightarrow\infty$, its round-trip reflection phase approaches $d\phi_{11}^C/d\omega\rightarrow d/c$. 
\bigskip{}

From the last two examples, note that, although a conjugate-mode pair gives $C\approx -I$ in the large-$\Gamma$ limit, when considering many such modes, their small pole contributions (deviations from $-I$), $\propto  i2\omega/\Gamma_\mathrm{o}$ from Eq.~(\ref{eq:C-conjugate-pair}), add up to give a non-trivial phase term (in transmission or reflection).

\subsection{Choice of boundary for $D$ calculation and $C$-matrix description of a port shift}
\label{C-from-port-shift}

\begin{figure}
\includegraphics[width=1\columnwidth,keepaspectratio]{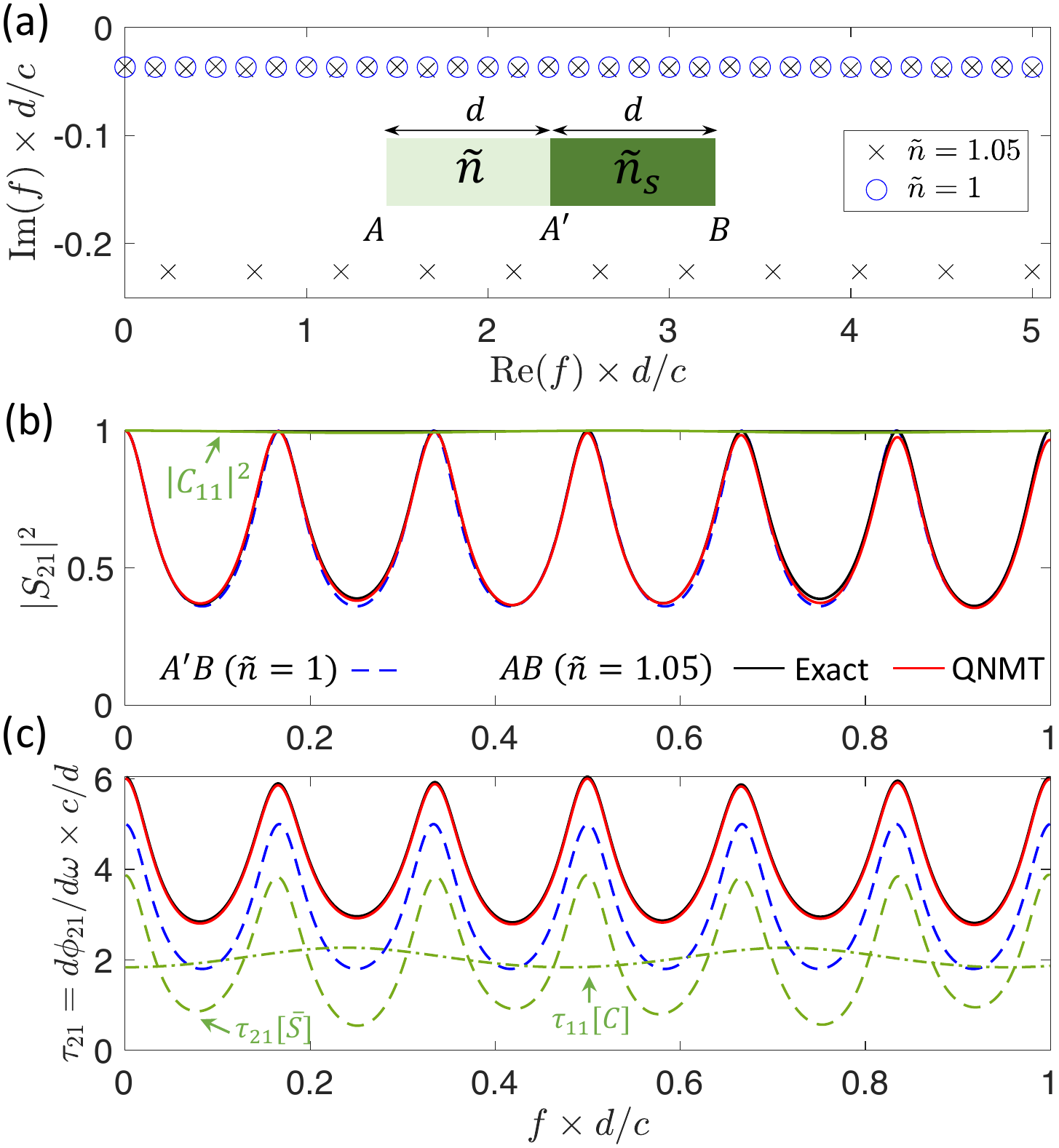}
\caption{Normally incident planewave transmission through slabs with indices $\tilde{n}_s=3$, $\tilde{n}=1.05$, and thickness $d$ (black for exact result and red for QNMT prediction), and comparison to the limiting case of $\tilde{n}=1$ (blue for exact result). (a) Geometry and poles, (b) amplitude and (c) group delay: the small shift is due to the extra propagation through the A--A$'$ slab, predicted by QNMT primarily via the contribution of the low-$Q$ modes, as can be seen also from the $\bar{S}C$ approximation (green) where $S_{21}\approx\bar{S}_{21}C_{11}$ gives the correct delay $\tau_{21}\approx\tau_{21}[\bar{S}]+\tau_{11}[C]$.}
\label{Fig-Dboundary} 
\end{figure}

In our QNMT formulation, we suggest calculating the coupling coefficient $D_{pn}$ as an overlap between the QNM~$n$ and the CPM~$p$ at the latter's cross-section $z'_p$ which first touches the scatterer boundary [Eq.~(\ref{D-definition})]. A reasonable question is whether such a choice is always appropriate, especially in unusual geometries where a thin ``needle'' sticks out from the scatterer or when very-low-index materials surround the scatterer. In such scenarios, the QNMs are expected to be localized close to the center of the scatterer and likely are already exponentially increasing inside the suggested outermost boundary. Here, we study a simple such photonic structure to explain the physics that come into play to render our boundary choice indeed suitable and we show how the $S=\bar{S}C$ formulation can give an interesting physical interpretation.

A plane wave at frequency $f=\omega/2\pi$ is normally incident on a uniform material slab of refractive index $\tilde{n}_s=3$ and thickness $d$ that is attached to another slab of the same thickness but with $\tilde{n}=1.05$ (\figref{Fig-Dboundary} inset). As a reference, the limiting symmetric system with $\tilde{n}=1$ and ports' cross sections A$'$, B on the boundaries of the $\tilde{n}_s$ slab has the response shown in \figref{Fig-Dboundary} for (a) the poles (blue circles) with $\sigma_n=\pm1$, (b) transmission amplitude, and (c) phase delay (blue dashed lines). The response of the actual test system (shown with black ``x'' and lines) obviously approaches that of the limiting case in amplitude and has an additional phase delay due to the propagation through the $\tilde{n}$ slab A--A$'$ [Figs.~\ref{Fig-Dboundary} (b) and \ref{Fig-Dboundary}(c)]. It may be tempting to think that the two systems should have the same QNMT parameters $(\omega_n,\sigma_n)$. However, its high-$Q$ modes now have $\sigma_n$ different from $\pm1$ (e.g., $\omega_n d /2\pi c=0.165-0.039i$ has $\sigma_n=-0.45+0.68i$), exactly because they are calculated on our suggested boundary $z'_p=$ A and the structure between A--B is asymmetric. How can we then expect QNMT to give an accurate prediction? The key lies in a set of very-low-$Q$ modes $\omega_n^C$ that are supported mainly by the weakly scattering $\tilde{n}$ slab [\figref{Fig-Dboundary}(a) below the inset] and have highly asymmetric $|\sigma_n^C|\ll1$ [e.g., $\omega_n^C d /2\pi c= 0.24-0.23i$ has $\sigma_n^C\sim10^{-3}(1+1i)$]. As $\tilde{n}\rightarrow1$, their $\Gamma_n^C\rightarrow\infty$ and, when taken into account, the QNMT prediction of $S$ gives the correct result with the additional expected group delay through A--A$'$.

To clearly understand the effect of the low-$Q$ modes, we separate them into a background response $C$, shown in green lines in \figref{Fig-Dboundary}. We see that $C$ is almost a diagonal matrix [$|C_{11}|^2\approx 1$ in \figref{Fig-Dboundary}(b)] and that $C_{11}$ represents a group delay equal to $2d/c$ [\figref{Fig-Dboundary}(c)]. Indeed, this spectrum of low-$Q$-modes matches the ``one-sided equi-spaced modes'' of the previous subsection: due to $n_s\gg n$, $A'$ acts as a highly reflective boundary, so $\tau_{11}^C$ models the roundtrip phase propagation through $A\rightarrow A'\rightarrow A$. The group delay of $\bar{S}_{21}$ through A--B is surprisingly reduced (and not increased!) by $d/c$ compared to the A$'$--B $S_{21}$, and that is because the high-$Q$ modes have modified $\sigma_n$ values. In this way, $S_{21}\approx \bar{S}_{21}C_{11}$ works out correctly to give the required additional group delay $d/c$. It is worth reminding that $\bar{S}$ does not have to be symmetric. In fact, oppositely to $\bar{S}_{21}$, the A--B $\bar{S}_{12}$ has group delay \emph{increased} by $d/c$ compared to the A$'$--B transmission, so that, combined with $C_{22}\approx-1\neq -e^{i\omega 2d/c}\approx C_{11}$, we get the correct $\bar{S}_{12}C_{22}\approx S_{12}=S_{21}\approx \bar{S}_{21}C_{11}$.

We conclude that all structural features contribute to scattering, which may be expressed via low-$Q$ modes that need to be included in the QNM expansion. This justifies our choice of surface for the calculation of $D$ in Eq.~(\ref{D-definition}) to be the closest port cross-section that encloses the entire scatterer $z'_p$. For a different port choice $z_p$, additional (practically impossible to locate) infinite-$\Gamma$ modes would need to be accounted for to express the extra phase shift through $z_p-z'_p$.

\subsection{Examples of Section \ref{sec-examples}}

We finally test this formulation on the 2-port electromagnetic structures of \figsref{Fig-test-structure} and \ref{Fig-angle} by using the modes with lowest $Q$s (much lower than the other modes) to construct the $C$ matrix (see modes in \figref{Fig-test-structure}(c) and the \SupMat). $C_{21}$ indeed yields a slowly-varying transmission background, as shown in \figsref{Fig-test-structure}(a) and \ref{Fig-angle}(a) (green dash-dotted lines). Sharp features in $S_{21}$ are then obtained by the high-$Q$ modes within the $\bar{S}C$ formulation (green solid lines), which actually gives a very good approximation of the overall transmission amplitude spectrum. Due to the remaining asymmetry of $\bar{S}C$, we find that there are some errors in the group-delay prediction close to transmission zeros, but these errors are reduced when considering realistic finite-bandwidth pulses, as shown in Appendix~\ref{App:group-delay}.

The 4-port structure of \figref{Fig-4-port} has large connected metallic sheets, so it exhibits some very-low-$Q$ ``one-sided equi-spaced modes" (\secref{sec_infinite-Gamma}) due to the outermost dielectric layers that only couple ports on the same side (namely they have $\sigma_{r,pq}\rightarrow 0$ or $\infty$ for $p,q$ in opposite sides, so that $|C_{pq}|\rightarrow 0$). This explains the sharp features observed in $S_{41}$, as only high-$Q$ modes contribute to it. Moreover, the layers in this example are very thin, so these low-$Q$ modes are located very far from the frequencies of interest anyway.

\section{Conclusions}

We have presented an expansion of the system scattering matrix $S$ over non-normalized QNMs, formulated to satisfy the fundamental physical conditions of realness, energy conservation and reciprocity even for a small truncated number of terms. Resonant QNMs with frequencies $\omega_n$ are computed with a numerical eigensolver. Coupling coefficients $D_{pn}$ are evaluated as surface overlap integrals between normalized-CPM and non-normalized-QNM fields (as only ratios $\sigma_{r,pn}=D_{pn}/D_{rn}$ are needed). Negative-frequency modes ($-\omega_n^{*},D_{pn}^{*}$) are included. The $D$ matrix is then adjusted through Eq.~(\ref{optimD}) and $S$ is finally calculated via Eq.~(\ref{phen_cmt}). For applications where it is convenient to separate an effective background response from the high-$Q$ resonances, $C$ can be determined by the same procedure using only the low-$Q$ resonances, with $\bar{S}$ from the high-$Q$ modes, and then $S \approx \bar{S}C$. In \secref{sec_infinite-Gamma}, we discussed several limiting cases, and showed that a nearly frequency-independent $C$ with nonzero transmission can be produced by a very-low-$Q$ mode on the imaginary-frequency axis, while a propagation phase is modeled by a set of several low-$Q$ modes. The agreement of our QNMT with exact simulations gives us confidence that it can be successfully employed for rapid device design. In a separate paper~\cite{DESIGN}, we indeed use this formulation to design precise standard (especially elliptic) high-order filters.

Our QNMT was mainly developed for linear ports with frequency-independent transverse mode profiles, such as plane waves. However, it could also be extended to finite arbitrary-shape scatterers using a spherical CPM basis. (Note that systems with spherical symmetry studied in previous QNMT formulations~\cite{alpeggiani2017quasinormal, zhang2020quasinormal} can be modeled merely as multiple 1-ports, so QNMT was not really needed, as explained in the introduction.) Difficulties arise with other types of ports, such as when QNM-to-CPM coupling coefficients $D(\omega)$ are frequency dependent, and specifically when the $S$ matrix has branch points due to CPM cutoff frequencies. While a rigorous extension of the theory to such systems may require a different approach, our model may still provide good approximate results for slowly varying coupling coefficients (such as for dielectric waveguides with low waveguide dispersion). 

\begin{acknowledgments}
This work was supported in part by the U.S. Army Research Office through the Institute for Soldier Nanotechnologies at MIT under Award No. W911NF-18-2-0048, by the Simons Foundation collaboration on Extreme Wave Phenomena, and by Lockheed Martin Corporation under Award No. RPP2016-005.
\end{acknowledgments}

\begin{appendix}

\section{Weierstrass factorization of $S$-matrix}
\label{app_weierstrass}

The Weierstrass factorization theorem~\cite{markushevich2013theory} states that a ``meromorphic'' function (analytic except for poles) can be factorized into a non-zero analytic function (an exponential) and a rational function (zeros and poles). If $S(\omega)$ is meromorphic, an exponential phase factor can then be factored out of each term as $S_{pq}(\omega)=e^{i\varphi_{pq}(\omega)}S'_{pq}(\omega)$, with $S'_{pq}$ a ``proper'' rational function (degree of numerator polynomial not larger than degree of denominator, so finite as $\omega\rightarrow\infty$) and $\varphi_{pq}$ an analytic function that we assume corresponds to a real phase shift for real frequencies. Similarly to the approach in \citeasnoun{nussenzveig1972causality}, the combination of Phragmén-Lindelöf theorem (giving $|e^{i\varphi_{pq}(\omega)}|\leq 1$ in the upper-half complex plane) and Čebotarev theorem shows that $\varphi_{pq}$ has to be a linear ($=\tau_{pq}\omega$ with $\tau_{pq} \geq 0$, since the constant term can be added to $S'_{pq}$).

When the system is lossless and reciprocal, unitarity and symmetry of $S$ combine to $S^*(\omega)S(\omega)=I$. The off-diagonal $pq$ term can be expanded as $\sum_r S^*_{pr}S_{rq}=\sum_r e^{-i\left(\tau_{pr}-\tau_{rq}\right)\omega}S'^*_{pr}S'_{rq}=0$. Since this has to be true for all real $\omega$, all the phase terms in the sum must be the same, namely $\tau_{pr}-\tau_{rq}=\theta_{pq}$ for each $r$. Specifically, for $r=p$ and $r=q$, we get $\tau_{pp}-\tau_{pq}=\tau_{pq}-\tau_{qq}\Leftrightarrow\tau_{pq}=\left(\tau_{pp}+\tau_{qq}\right)/2$. Therefore, the $S$ matrix can be written as $S=e^{i\tau \omega}S'e^{i\tau \omega}$ with a \emph{diagonal} $\tau$ matrix with real positive elements.

\section{Energy conservation condition Eq.~(\ref{K-for-unitarity})}
\label{app_unitarity}

For a lossless system, $S$ is unitary. From Eq.~(\ref{S-expansion}),
namely $S=-I-DdK^{t}$, where $d=(i\omega-i\Omega)^{-1}$:
\begin{equation}
S^{\dagger}S=I+K^{*}d^{\dagger}D^{\dagger}+DdK^{t}+K^{*}d^{\dagger}D^{\dagger}DdK^{t}.
\end{equation}
We compute the coefficient $(p,q)$ of this matrix. The second and third terms are equal to: 
\begin{equation}
\sum_{n}\frac{K_{pn}^{*}D_{qn}^{*}}{i(\omega-\omega_{n}^{*})}-\sum_{n}\frac{D_{pn}K_{qn}}{i(\omega-\omega_{n})}.
\end{equation}
The final term, after decomposing into simple elements through $\frac{1}{i(\omega-\omega_{n})i(\omega-\omega_{l}^{*})}=\frac{1}{i(\omega_{n}-\omega_{l}^{*})}[\frac{1}{i(\omega-\omega_{n})}-\frac{1}{i(\omega-\omega_{l}^{*})}]$
and relabelling indices ($n$ and $l$), becomes: 
\begin{equation}
\begin{split} 
& \sum_{r,n,l}\frac{K_{pl}^{*}D_{rl}^{*}D_{rn}K_{qn}}{i(\omega-\omega_{l}^{*})i(\omega-\omega_{n})}=\\
= & \sum_{n}\frac{K_{qn}}{i(\omega-\omega_{n})}\sum_{l}K_{pl}^{*}\frac{\sum_{r}D_{rn}D_{rl}^{*}}{i(\omega_{n}-\omega_{l}^{*})}\\
- & \sum_{n}\frac{K_{pn}^{*}}{i(\omega-\omega_{n}^{*})}\sum_{l}K_{ql}\frac{\sum_{r}D_{rl}D_{rn}^{*}}{i(\omega_{l}-\omega_{n}^{*})}.
\end{split}
\end{equation}
In order to impose $S^{\dagger}S=I$ for every $\omega$, necessary and sufficient conditions are given by $K^{*}M=D$ and $KM^{t}=D^{*}$, with $M_{nl}=\frac{\sum_{p}D_{pl}D_{pn}^{*}}{i(\omega_{l}-\omega_{n}^{*})}$. The two relations are actually equivalent, since $M=M^{\dagger}$, and can be rewritten as $K^{t}=M^{-1}D^{\dagger}$.

\section{Realness under Eq.~(\ref{K-for-unitarity})}
\label{app_real_fields}

Including negative-frequency modes with $D_{pn'}=D_{pn}^{*}$, we can write $M=\begin{pmatrix}A & B\\B^* & A^*\end{pmatrix}$ (where $A=A^{\dagger}$ and $B=B^{t}$, so that $M=M^{\dagger}$). Now, consider matrices $\tilde{A}$, $\tilde{B}$ such that $\begin{pmatrix}\tilde{A} & \tilde{B}\end{pmatrix}M=\begin{pmatrix}I & 0\end{pmatrix}$; then, by directly substituting $M$, we can immediately show that we also have $\begin{pmatrix}\tilde{B}^{*} & \tilde{A}^{*}\end{pmatrix}M=\begin{pmatrix}0 & I\end{pmatrix}$, so $M^{-1}=\begin{pmatrix}\tilde{A} & \tilde{B}\\ \tilde{B}^{*} & \tilde{A}^{*} \end{pmatrix}$. Finally, from $K=D^{*}\left(M^{t}\right)^{-1}$, we conclude that $K_{pn'}=K_{pn}^{*}$.

\section{Normalization independence of Eq.(\ref{phen_cmt})}
\label{app_uniqueness} 

Let $D_{r}$ be a $N\times N$ diagonal matrix, with elements $D_{r,nn}=D_{r_{n}n}$, for some chosen port $r_{n}$ for each resonant mode $n$ (usually, we simply take $r_{n}=1$). Then, denote by $\sigma_{r}=DD_{r}^{-1}$ the $r$-scaled $P\times N$ coupling matrix (naturally, $\sigma_{r,r_{n}n}=1$ for all $n$). Now, by inserting $I$ twice in Eq.~(\ref{phen_cmt}), the scattering matrix $S$ from can be rewritten as 
\begin{align}\label{eq:S-normalized}
S & =-I-D(i\omega-i\Omega)^{-1}M^{-1}D^{\dagger}\nonumber \\
 & =-I-D\left(D_{r}^{-1}D_{r}\right)(i\omega-i\Omega)^{-1}M^{-1}\left[D_{r}^{\dagger}\left(D_{r}^{-1}\right)^{\dagger}\right]D^{\dagger}\nonumber \\
 & =-I-\sigma_{r}(i\omega-i\Omega)^{-1}M_{r}^{-1}\sigma_{r}^{\dagger},
\end{align}
where
\begin{equation}
\begin{split}
&M_{r}=\left(D_{r}^{-1}\right)^{\dagger}MD_{r}^{-1}\Leftrightarrow \\
M_{r,nl}=\frac{1}{D_{r_{n}n}^{*}}&\frac{\sum_{p}D_{pl}D_{pn}^{*}}{i\omega_{l}-i\omega_{n}^{*}}\frac{1}{D_{r_{l}l}}=\frac{1+\sum_{p\neq r_{n}}\sigma_{r,pl}\sigma_{r,pn}^{*}}{i\omega_{l}-i\omega_{n}^{*}}.
\end{split}
\label{eq:M-normalized}
\end{equation}
These two equations show that, for a lossless system, $S$ in Eq.~(\ref{phen_cmt}) can be fully computed using only the resonant frequencies $\Omega$ and the ratios $\sigma_{r}$ of modal coupling among different ports, independently of the overall scaling factors in $D$.

For 2-port systems, we simply choose $r_n=1$, skip the $r$ subscript and denote $\sigma_{n}=D_{2n}/D_{1n}$. The $S$ matrix then becomes
($p,q=1,2$)
\begin{align}
 & S_{pq}=-\delta_{pq}-\sum_{n=1}^{N}\frac{\sigma_{pn}\sum_{l=1}^{N}M^{-1}_{\;nl}\sigma_{ql}^{*}}{i\omega-i\omega_{n}}\label{eq:2-port-S21}\\
 & M_{nl}=\frac{1+\sigma_{l}\sigma_{n}^{*}}{i\omega_{l}-i\omega_{n}^{*}};\;\sigma_{1n}=1,\sigma_{2n}=\sigma_{n}.\nonumber 
\end{align}
When the system has mirror symmetry, these ratios can only take the values $\sigma_{n}=\pm1$, so they act like eigenvalues of the symmetry operator.

Reversely, we also show that $S$ uniquely determines $\Omega$ and $\sigma_{r}$. In particular, for two different sets $\{\Omega,\sigma_{r}\}$, $\{\Omega',\sigma'_{r}\}$ such that  $S_{\{\Omega,\sigma_{r}\}} = S_{\{\Omega',\sigma'_{r}\}}$, we see from Eq.~(\ref{eq:S-normalized}) that $\Omega=\Omega'$ and
\begin{equation}
    \sigma_{r,pn}\left[\sigma_rM_r^{-1}\right]_{qn}^* = \sigma'_{r,pn} \left[\sigma'_rM'^{-1}_r\right]_{qn}^*.\label{eq:C4}
\end{equation}
For $p=r_n$, we have $\left[\sigma_rM_r^{-1}\right]_{qn} = \left[\sigma'_rM'^{-1}_r\right]_{qn}$. Then, for any $p$, Eq.~(\ref{eq:C4}) again gives $\sigma_{r,pn} = \sigma'_{r,pn}$.

\section{Properties of zeros of lossless reciprocal 2-port systems}
\label{app_zeros-2ports} 

Combining realness $S^{*}(\omega)=S(-\omega)$, unitarity $S^{\dagger}(\omega)S(\omega)=I$ and symmetry $S^{t}(\omega)=S(\omega)$ on the real-$\omega$ axis, gives $S(-\omega)S(\omega)=I$, which can then be analytically continued in the entire complex-$\omega$ plane. Expanding this equation for a 2-port, we get 
\begin{subequations}
\begin{align}
S_{11}\left(-\omega\right)S_{11}\left(\omega\right)+S_{21}\left(-\omega\right)S_{21}\left(\omega\right)=1\label{eq:S-properties-a}\\
S_{11}\left(-\omega\right)S_{11}\left(\omega\right)=S_{22}\left(-\omega\right)S_{22}\left(\omega\right)\label{eq:S-properties-c}\\
S_{11}\left(-\omega\right)S_{21}\left(\omega\right)=-S_{21}\left(-\omega\right)S_{22}\left(\omega\right).\label{eq:S-properties-b}
\end{align}
\end{subequations}
We can use these equations to derive some useful properties of the zeros of $S$ coefficients. As a reminder, realness requires all poles and zeros to be symmetric across the imaginary $\omega$ axis. 

Let us first assume that $-\omega_{o}$ is \emph{not} a system pole [so $S(-\omega_\mathrm{o})$ is finite]. Eq.~(\ref{eq:S-properties-a}) then prevents $S_{11}$ and $S_{21}=S_{12}$ from having simultaneous zeros at $\omega_{o}$, and using also Eq.~(\ref{eq:S-properties-c}) the same holds for $S_{22}$ and $S_{21}$. Therefore,  Eq.~(\ref{eq:S-properties-b}) mandates that (i) the zeros of $S_{21}$ can only appear as complex quadruplets  ($\omega_{o},\omega_{o}^{*},-\omega_{o},-\omega_{o}^{*}$), real or imaginary pairs ($\omega_{o},-\omega_{o}$), or at $\omega_{o}=0$ and that (ii), for each zero-pair ($\omega_{o},-\omega_{o}^{*}$) of $S_{11}$, ($-\omega_{o},\omega_{o}^{*}$) is a zero pair of $S_{22}$.

If now $-\omega_{o}$ is a system pole (so at least one $S$-element diverges there), the same rules (i) and (ii) still apply, as long as zero-pole cancellations that occur at ($-\omega_{o},\omega_{o}^{*}$) are taken into account. All possible scenarios are: (I) If $S_{11}\left(\omega_{o}\right)=0\neq S_{22}\left(\omega_{o}\right)$, Eq.~(\ref{eq:S-properties-c}){} forces $S_{22}\left(-\omega_{o}\right)$ to be finite, thus $S_{22}$ must exhibit a zero at $-\omega_{o}$ that cancels the pole there (and similarly when switching $S_{11}$ and $S_{22}$). Moreover, if simultaneously $S_{21}\left(\omega_{o}\right)=0$, Eq.~(\ref{eq:S-properties-b}){} mandates a pole-zero cancellation also for $S_{21}$ at $-\omega_{o}$. (II) If $S_{11}\left(\omega_{o}\right)=S_{22}\left(\omega_{o}\right)=0$, Eq.~(\ref{eq:S-properties-b}) implies that also $S_{21}\left(\omega_{o}\right)=0$, namely the entire matrix $S(\omega_\mathrm{o})=0$. In this case, we can consider that all $S$-coefficients also have another zero at $-\omega_{o}$ that cancels a degenerate pole, which usually do appear when the physical system is perturbed. (III) If $S_{11}\left(\omega_{o}\right)\neq0\neq S_{22}\left(\omega_{o}\right)$, Eqs.~(\ref{eq:S-properties-a}) and (\ref{eq:S-properties-b}) dictate also that $S_{21}\left(\omega_{o}\right)\neq0$.  In the cases above, for any coefficient with $S_{pq}(\omega_\mathrm{o})=0$, then $S_{pq}(\omega)\propto e^{i\varphi(\omega)}= \left(\omega-\omega_{o}\right)\left(\omega+\omega_{o}^{*}\right)/\left(\omega+\omega_{o}\right)\left(\omega-\omega_{o}^{*}\right)$, a behavior known as an ``all-pass'' phase filter. Moreover, when $S(\omega_\mathrm{o})$ is singular, $\omega_\mathrm{o}$ is called an ``$S$-matrix zero'' (for example generalized in~\citeasnoun{sweeney2020theory}).

The restrictions (i) on the $S_{21}$ zeros imply that its numerator is a polynomial of $\omega^{2}$ with real coefficients, optionally with multiplicative $i\omega$ factors.

\section{Guidelines for numerical calculation of eigenmodes}
\label{App:finite-element}

Our QNMT relies on the calculation of the system resonances, including low-$Q$ modes, which may be spread across the complex-$\omega$ plane and may include a mode with zero real frequency ($\omega_\mathrm{o}=0-i\Gamma_\mathrm{o}$)---required for nonzero background transmission. Accurate numerical solution of the exact field equations for such eigenmodes can be challenging due to issues arising from spatial discretization and from truncation of the simulation domain in open systems, usually implemented with a ``perfectly matched layer'' (PML)~\cite{taflove2005computational, Johnson2010pml}. In particular, a PML introduces the so-called ``PML modes''~\cite{yan2018rigorous, vial2014quasimodal}, which can pollute the complex-$\omega$ plane and thus make it difficult to find the system's low-$Q$ QNMs. Here, we suggest some simulation-parameter choices that seem to help calculate QNMs accurately, including the low-$Q$ ones. For subwavelength structures (like the metallic microwave metasurfaces considered here), where quasi-static resonances can be sensitive to small spatial features, finite-element methods are attractive because they enable nonuniform meshing, so we used COMSOL Multiphysics~\citep{comsol}.\\

\paragraph*{Mesh type---}
In the computational domain (the cladding) around the scatterer (and in the PML when present), to minimize spurious reflections from changes in discretization one should use a mesh that is periodic (or nearly so) in the direction orthogonal to the boundary (where scattered waves should escape)~\cite{taflove2005computational, Johnson2010pml}. In COMSOL, we employed a ``swept'' mesh for this reason, and found it to give smoother convergence of the complex eigenfrequencies with mesh density compared to a standard tetrahedral mesh and to work better for modes with zero real frequency.\\

\begin{figure}
\includegraphics[width=1\columnwidth,keepaspectratio]{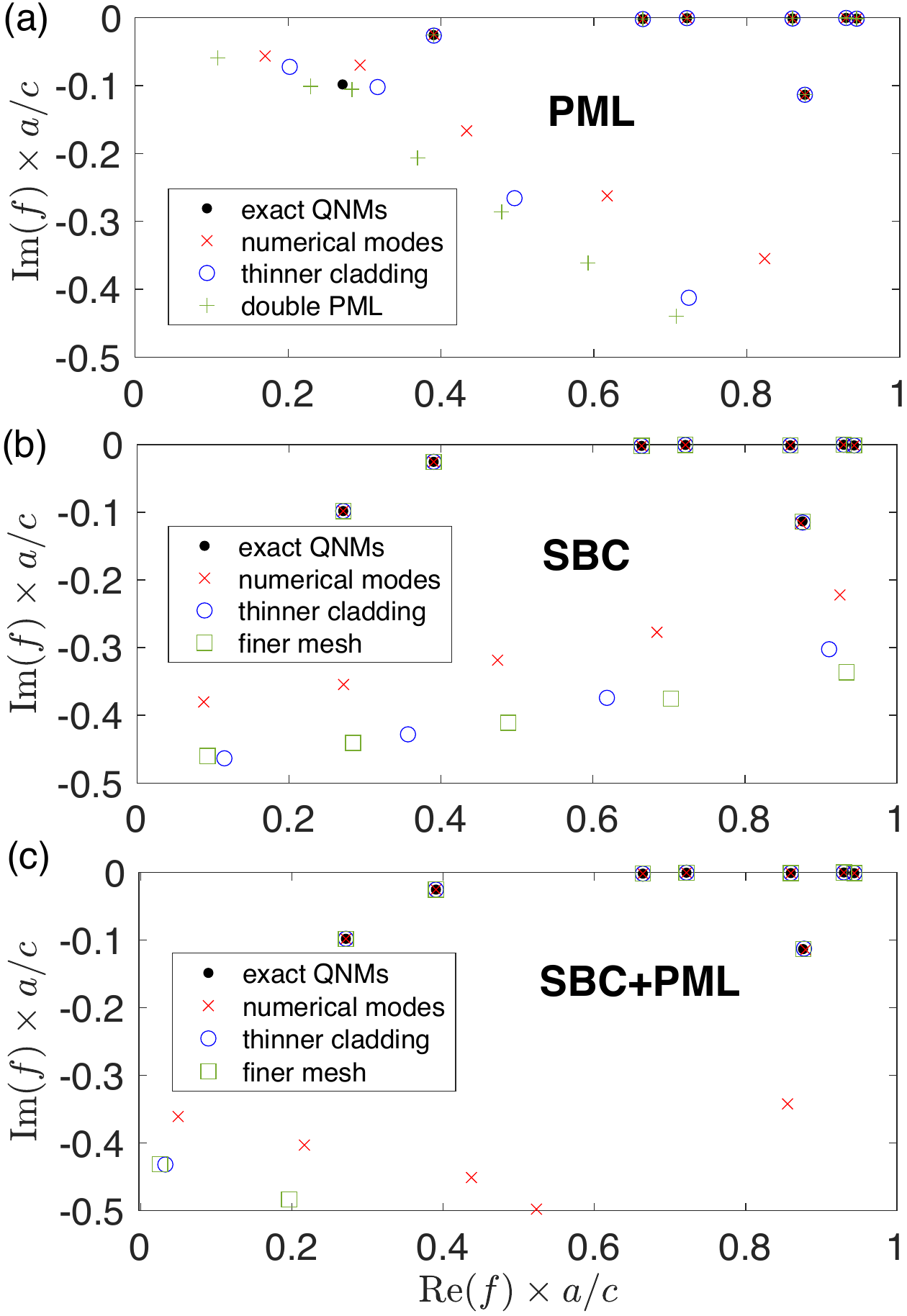} 
\caption{Modes computed numerically for a metasurface (of period $a=\,15 mm$) with cross-like apertures (of slits' width $0.02a$ and length $0.245a$) on a metallic sheet sandwiched between two dielectric layers (of permittivity $\epsilon=3$ and thickness $0.5a$). We only calculate the normally radiating modes of even symmetry, using cladding thickness $t_{\textrm{c}}$, PML thickness $t_{\textrm{PML}}$, mesh element size $h$ in the cladding/PML, and: (a) PML backed by a perfect electric conductor (``x'': $t_{\textrm{c}}=t_{\textrm{PML}}=0.5a$, $h=a/10$; ``o'': $t_{\textrm{c}}=0.25a$; ``+'': $t_{\textrm{PML}}=a$), (b) SBC without PML (``x'': $t_{\textrm{clad}}=a$, $h=a/6$; ``o'': $t_{\textrm{clad}}=0.75a$; ``+'': $h=a/10$), and (c) SBC with PML (``x'': $t_{\textrm{c}} = t_{\textrm{PML}}=0.5a$, $h=a/6$; ``o'': $t_{\textrm{clad}}=0.25a$; ``+'': $h=a/10$). The low-$Q$ QNM at $0.27-0.1i$ is not computed accurately when using only a PML due to the nearby radiation modes, while those are pushed down in the complex plane when using a SBC.}
\label{Fig-pml-modes}
\end{figure}

\paragraph*{Boundary condition, computational-cell size and mesh density---}
The commonly used frequency-independent PML of finite thickness rotates the continuous radiation spectrum in the complex plane and discretizes it into so-called ``PML modes''~\cite{yan2018rigorous, vial2014quasimodal}. Only QNMs above the rotated continuum can be accurately computed. The rotation is larger for a more absorptive (stronger or thicker) PML and for a thinner cladding, while the mesh density has little effect for fixed PML absorption. We illustrate these points for a metasurface example in~\figref{Fig-pml-modes}(a): the exact QNMs are indicated with black dots, while the other symbols mark all the modes obtained numerically for different sets of computational parameters. Note that the low-$Q$ QNM at $\approx0.27-0.1i$ cannot be calculated accurately, unless one used an even larger rotation that could be computationally expensive. Moreover, it turns out that the field profiles of the QNM and nearby PML modes are quite similar, yet another obstacle to distinguishing the former. Even worse, for structures supporting a zero-frequency QNM ($-i\Gamma_\mathrm{o}$), this mode cannot be found with a frequency-independent PML, as the negative imaginary axis will always lie below any rotated radiation spectrum~\cite{nannen2016spurious}.

Our QNMT is mostly applicable to CPMs with a frequency-independent transverse field profile. Thus, while computing QNMs, a ``scattering boundary condition'' (SBC) matching each CPM's profile can be used. As we show in~\figref{Fig-pml-modes}(b), using an SBC but not a PML, the discrete radiation modes are now simply pushed down in the complex plane (instead of rotated as with the PML, although note that a similar push-down occurs with a frequency-\emph{dependent} PML~\cite{nannen2016spurious}). This clears the complex plane (up to the first diffraction branch point) and allows for an accurate computation of the low-$Q$ and zero-frequency QNMs. The modal push is deeper for a thinner cladding and a denser mesh: these SBC-based discrete radiation modes disappear in the limit of infinite resolution, in contrast to PML modes~\cite{nannen2016spurious}.

Combining a PML and a SBC gives an even better result, as the discrete radiation modes are pushed down and then rotated in the complex plane [\figref{Fig-pml-modes}(c)]. In summary, when possible, it is advisable to use SBCs matching the CPMs, along with a densely meshed PML and a thin cladding (which can practically be kept at a minimum for a PML mesh fine enough to also adequately model the QNM near fields).\\

\paragraph*{Initial guesses for QNMs---}
Even if the discrete radiation modes have been pushed away, locating the low-$Q$ (and zero-frequency) QNMs initially may require good initial complex-frequency guesses and searching for a large number of modes. In some convenient cases, the initial guesses can be provided by the analytically-solveable modes of an effective average structure. We discussed such cases in paragraphs ``Equi-spaced (Fabry--Perot) modes'' and ``One-sided equi-spaced modes'' in \secref{sec_infinite-Gamma}.\\

\paragraph*{Number of QNMs needed---}
Regarding how many modes one needs in QNMT for an accurate enough calculation, one observation is that a (low-$Q$) mode $\omega_{n}=\Omega_{n}-i\Gamma_{n}$ will have an effect at real frequency $\omega$, if $\omega$ lies within a bandwidth proportional to the resonance linewidth, namely if $\left|\omega-\Omega_{n}\right|/\Gamma_{n}<{}$ some number, which depends on the level of accuracy required. Obviously, when a mode has very large $\Gamma_{n}$, it can play a role even if it is very distant and it may be very difficult to locate. In those cases, a background response $C$ can still be estimated via $C=\bar{S}^{-1}S$, with $S$ from a direct simulation and $\bar{S}$ from QNMT using only the high-$Q$ modes.

\section{Group delay for a finite-width pulse}
\label{App:group-delay}

For finite-bandwidth signal pulses, the group delay of a system should be computed as the difference in time expectation of flux between the input and output signals~\cite{peatross2000average}. In general, for a signal $f(t)$ with Fourier transform $F(\omega)=A_Fe^{i\phi_F}$ (amplitude $A_F$ and phase $\phi_F$), the flux time expectation is given by:
\begin{equation} \begin{gathered}
\langle t\rangle_f = \frac{\int t|f(t)|^2dt}{\int |f(t)|^2dt} \\
= \frac{-i\int F'(\omega)F^*(\omega)d\omega}{\int |F(\omega)|^2d\omega}=\frac{\int \phi'_F(\omega)|F(\omega)|^2d\omega}{\int |F(\omega)|^2d\omega},
\end{gathered} \end{equation}
where $'$ denotes a frequency derivative $d/d\omega$ and the integral involving $A_F'(\omega)$ cancels due to realness [$A_F(-\omega)=A_F(\omega)\Rightarrow A_F'(-\omega)=-A_F'(\omega)$].

\begin{figure}[!h]
\includegraphics[width=1\columnwidth,keepaspectratio]{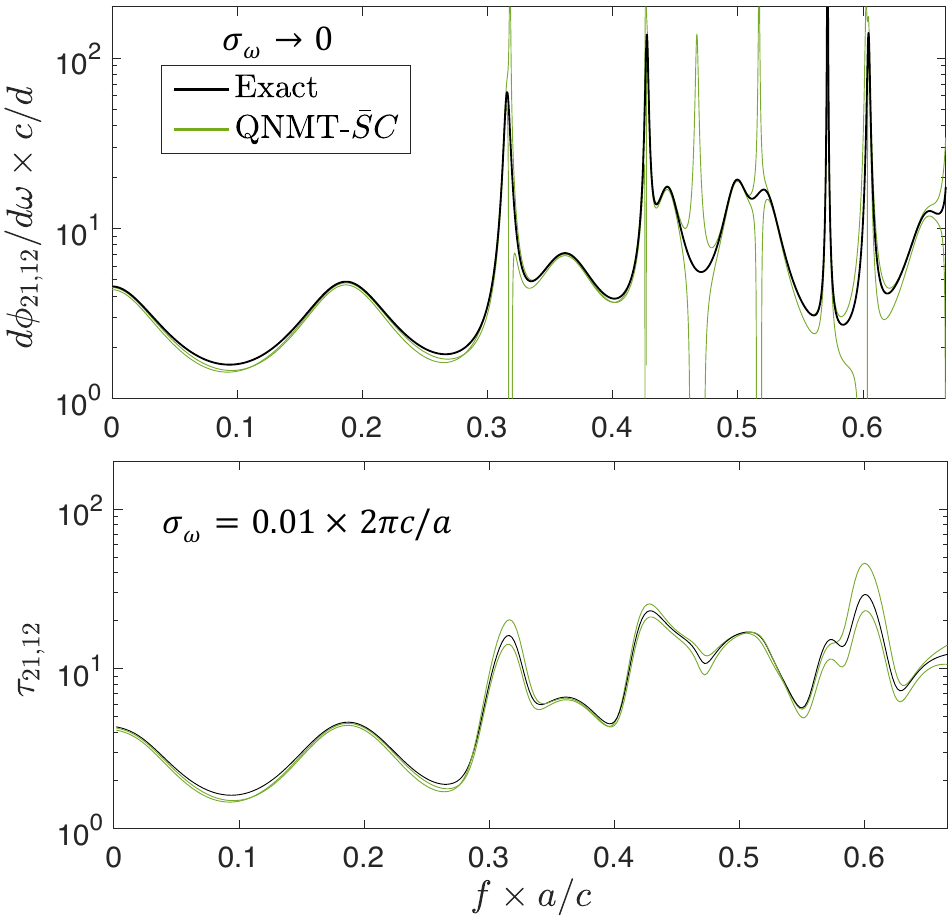} 
\caption{Group delay of $S_{21}$ and $S_{12}$ for oblique incidence on photonic grating presented in \figref{Fig-angle}. The asymmetric $S=\bar{S}C$ approximation (green) has errors for (a) $d\phi/d\omega$ but agrees much better with the exact result (black) for (b) a more realistic Gaussian input pulse of spectral width $\sigma_\omega=0.01\times 2\pi c/a$.}
\label{Fig-TimeDelay} 
\end{figure}

For a system with transfer function $H(\omega)$, the output signal $g(t)$ has Fourier transform $G(\omega) = H(\omega)F(\omega)$, so $\phi'_G(\omega)=\phi'_H(\omega)+\phi'_F(\omega)$. The system group delay is then equal to $\tau = \langle t\rangle_g -\langle t\rangle_f$ and, when $\phi'_F(\omega)$ is constant (e.g., unchirped Gaussian pulse), it simplifies to:
\begin{equation}
\tau = \frac{\int \phi'_H(\omega)|H(\omega)|^2|F(\omega)|^2d\omega}{|H(\omega)|^2|F(\omega)|^2d\omega}.
\label{eq:time-delay}
\end{equation}
In the limit of a very narrowband ($\delta$-function) input frequency spectrum $F$, this simplifies to the usual $\tau\rightarrow \phi'_H$.

In all QNMT formulations which do not simultaneously satisfy reciprocity and energy conservation (e.g., using Eq.~(7) without Eq.~(11), our $S=\bar{S}C$ formula, or Refs.~\cite{alpeggiani2017quasinormal, weiss2018qnmt}), the zeros of $S$ coefficients do not abide by the required relations in the complex plane and this leads to errors in the phase (group-delay) response. To visualize them, in \figref{Fig-TimeDelay}(a) we plot the ($\delta$-function) group delays $\phi'_H$ of $H=S_{21}$ and $H=S_{12}$ for our approximate non-reciprocal formula $S=\bar{S}C$ for the oblique-incidence example of Fig.~4. We see that they mostly coincide (among them and with the exact result), except for frequency regions where the exact $S_{21}$ goes to zero, and in which $\tau_{21}$, $\tau_{12}$ exhibit erroneous deviating spikes, often indicating ``superluminal'' ($<1$) or negative delays. However, Eq.~(\ref{eq:time-delay}) shows that frequencies corresponding to small amplitude $H(\omega)$ do not contribute much to the real time delay of a finite-bandwidth pulse. We therefore plot in \figref{Fig-TimeDelay}(b) again the exact and both $\bar{S}C$ delays for an input Gaussian pulse $F(\omega) \propto \textrm{exp}[-(\omega-\omega')^2/4\sigma_\omega^2]$ with $\sigma_\omega=0.01\times 2\pi c/a$. We see that the erroneous spikes have now disappeared and we get much better agreement, which improves as $\sigma_\omega$ increases, as we confirmed.
\end{appendix}

\bibliography{biblio}{}

\pagebreak
\widetext
\newpage
\begin{center}
\textbf{\large Supplemental Material}
\end{center}
\setcounter{equation}{0}
\setcounter{figure}{0}
\setcounter{table}{0}
\setcounter{page}{1}
\setcounter{section}{0}
\makeatletter
\renewcommand{\theequation}{S\arabic{equation}}
\renewcommand{\thefigure}{S\arabic{figure}}

\section{4-port metasurface via coupled polarizations example}

\begin{figure}[!htp]
\includegraphics[width=0.9\columnwidth,keepaspectratio]{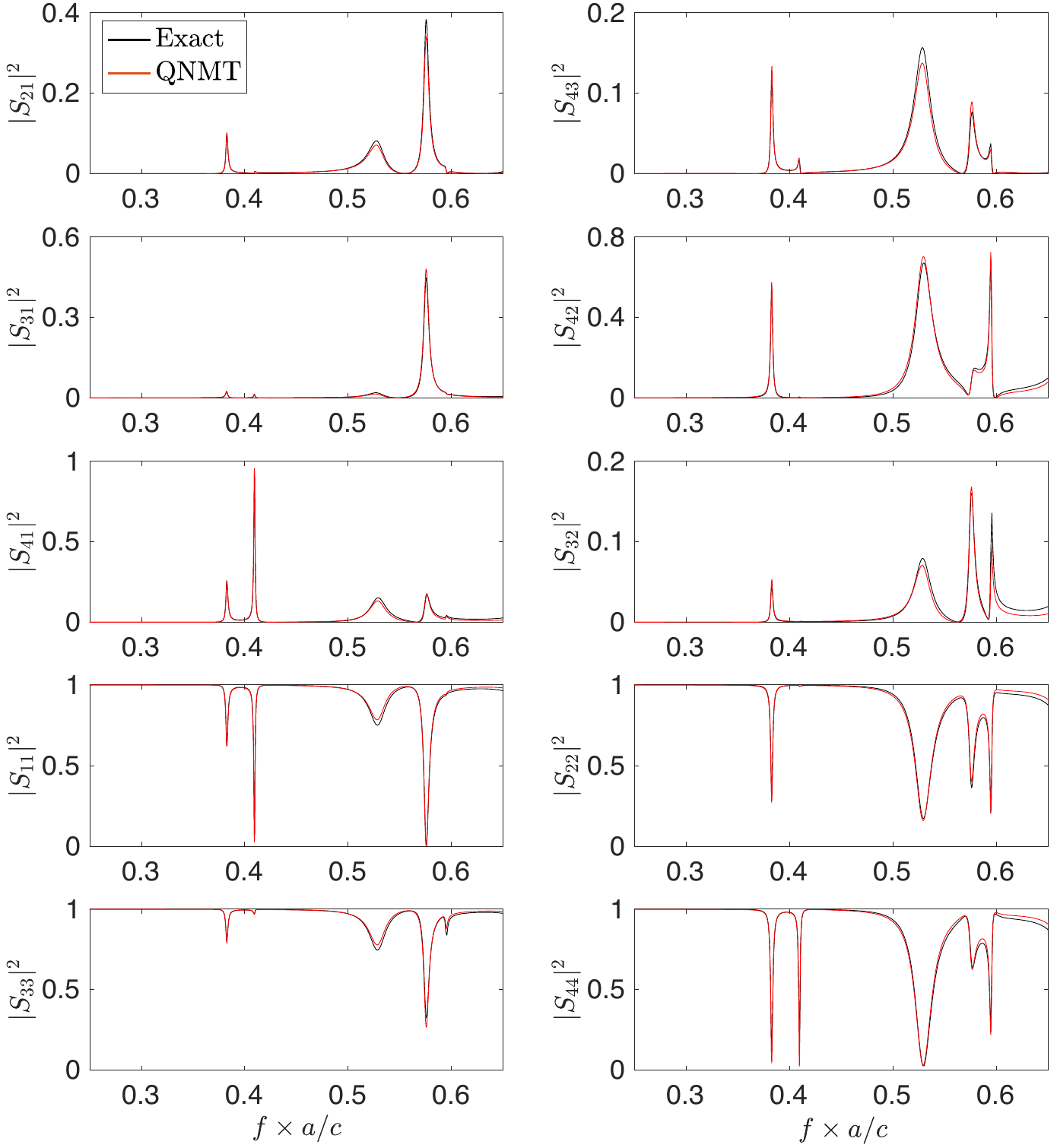} 
\caption{All components of $S$ for the 4-port metasurface presented in the main text Fig.~3. The vertical axis is scaled on each plot for better visibility. The agreement of QNMT (red) to the exact result (black) is very good for all coefficients.}
\label{Fig-4port} 
\end{figure}
\newpage

\section{QNM parameters for structures}

Here, we provide all the QNMs computed via finite-element simulations. The computed QNM-to-CPM ratios are indicated by $\sigma^c$, while the ones finetuned by Eq.~(11) of the main text are denoted by $\sigma$. The modes used to calculate the background $C$ matrix are marked in bold. For simplicity, $\Omega$ and $\Gamma$ in the tables denote the dimensionless frequency values as defined in the corresponding figures.\\

\textbullet\ Normal incidence on microwave metasurface of Fig.~2:

\begingroup
\fontsize{9}{12}\selectfont
\begin{center}
\begin{tabular}{ | c | c | c | c | c | c | c | c | c | c | c | c | } 
 \hline
 $\Omega$ & \textbf{0} & 0.2020 & 0.2929 & \textbf{0.3510} & \textbf{0.5099} & 0.6207 & 0.6431 & 0.6539 & 0.6932 & \textbf{0.7395} \\
 \hline
 $\Gamma$ & \textbf{0.0699} & 0.0136 & 0.0218 & \textbf{0.0625} & \textbf{0.0905} & 0.0029 & 0.0145 & 0.0048 & 0.0048 & \textbf{0.0363} \\
 \hline
 $\sigma^c$ &  \textbf{0.98} & -1.01+0.09i & 0.14+0.01i & \textbf{1.23+4.29i} & \textbf{-0.27+0.11i} & -0.81+0.18i & -0.07+0.19i & -0.26-1.70i & -0.93-0.36i & \textbf{0.89+0.2i} \\
 \hline
 $\sigma$ &  \textbf{1.00} & -1.04+0.07i & 0.14+0.02i & \textbf{1.14+4.50i} & \textbf{-0.28+0.09i} & -0.78+0.12i & -0.09+0.20i & -0.05-1.67i & -0.94-0.61i & \textbf{0.75+0.3i} \\
 \hline
 $\tilde{\Omega}$ & \textbf{0} & 0.2018 & 0.2925 & \textbf{0.3503} & \textbf{0.5091} & 0.6206  & 0.6428 & 0.6538 & 0.6931 & \textbf{0.7388} \\
  \hline
 $\tilde{\Gamma}$ & \textbf{0.0699} & 0.0147 & 0.0234 & \textbf{0.0641} & \textbf{0.0926} & 0.0056 & 0.0171 & 0.0077 & 0.0081 & \textbf{0.0397} \\
 \hline
\end{tabular}
\end{center}
\endgroup

\textbullet\ 4-port metasurface via coupled polarizations of Fig.~3:
\begin{center}
\begin{tabular}{ | c | c | c | c | c | c | c | } 
 \hline
 $\Omega_n$ & 0.3826 & 0.4093 & 0.5288 & 0.5753 & 0.5948 & 0.6753 \\
 \hline
 $\Gamma_n$ & 0.0011 & 0.0007 & 0.0095 & 0.0029 & 0.0011 & 0.0064 \\
  \hline
 $\sigma^c_{1,2n}$ &  -1.44-0.36i & -0.07+0.01i & 2.00+0.62i & 0.57+0.23i & -17.60-5.09i & 1.68+1.69i \\
 \hline
 $\sigma_{1,2n}$ &  -1.50-0.13i & -0.06+0.03i & 2.24-0.21i & 0.59-0.06i & -14.57+3.13i & 3.17-0.01i \\
   \hline
 $\sigma^c_{1,3n}$ &  0.71+0.00i & 0.15+0.03i & 1.01-0.02i & 0.66-0.03i & -3.62+2.09i & -0.71-0.03i \\
 \hline
 $\sigma_{1,3n}$ &  0.72+0.00i & 0.13+0.03i & 1.01-0.03i & 0.70-0.02i & -2.40+1.190i & -0.72-0.03i \\ 
 \hline
 $\sigma^c_{1,4n}$ &  -2.32-0.60i & 1.20+0.25i & 2.83+0.79i & 0.43+0.00i & 9.97+4.22i & -1.37-1.40i \\
 \hline
 $\sigma_{1,4n}$ &  -2.37-0.23i & 1.18+0.01i & 3.11-0.36i & 0.41-0.17i & 8.90-1.07i & -2.60+0.00i\\
  \hline
\end{tabular}
\end{center}

\textbullet\ Oblique incidence on 2d photonic metasurface of Fig.~4:

\begin{center}
\begin{tabular}{ | c | c | c | c | c | c | c | c | c | } 
 \hline
 $\Omega$ & \textbf{0} & \textbf{0.187} & 0.3157 & \textbf{0.362} & 0.427 & \textbf{0.444} & \textbf{0.499} \\
 \hline
 $\Gamma$ & \textbf{0.0402} & \textbf{0.0377} & 0.0026 & \textbf{0.0265} & 0.0012 & \textbf{0.0107} & \textbf{0.0106} \\
 \hline
 $\sigma^c$ &  \textbf{0.953} & \textbf{-0.679+0.03i} & 0.345+0.274i & \textbf{0.992-0.051i} & 1.08-0.181i & \textbf{-0.841+0.062i} & \textbf{-1.71+1.19i} \\
 \hline
 $\sigma$ &  \textbf{0.973} & \textbf{-0.689-0.019i} & 0.289+0.329i & \textbf{1.11+0.088i} & 1.21+0.032i & \textbf{-0.787-0.073i} & \textbf{-1.99+0.865i} \\
 \hline
 $\tilde{\Omega}$ & \textbf{0} & \textbf{0.184} & 0.3152 & \textbf{0.359} & 0.4264 & \textbf{0.443} & \textbf{0.498} \\
 \hline
 $\tilde{\Gamma}$ & \textbf{0.0402} & \textbf{0.0440} & 0.0121 & \textbf{0.0391} & 0.0180 & \textbf{0.0237} & \textbf{0.0274} \\
 \hline
\end{tabular}
\end{center}

\begin{center}
\begin{tabular}{ | c | c | c | c | c | c | c | c | } 
 \hline
 $\Omega$ & \textbf{0.5219} & 0.5716 & 0.6044 & \textbf{0.6514} & 0.6737 & 0.7231 \\
 \hline
 $\Gamma$ & \textbf{0.0122} & 0.00024 & 0.0011 & \textbf{0.016} & 0.0065 & 0.0052 \\
 \hline
 $\sigma^c$ & \textbf{-0.078+0.24i} & -0.051+0.137i & -3.09-0.498i & \textbf{0.409+0.012i} & 1.39+3.76i & 0.246-0.008i \\
 \hline
 $\sigma$ & \textbf{-0.125+0.221i} & -0.081+0.120i & -2.96-1.43i & \textbf{0.369+0.155i} & 0.769+5.40i & 0.266-0.024i \\
 \hline
 $\tilde{\Omega}$ & \textbf{0.5195} & 0.5708 & 0.6032 & \textbf{0.6499} & 0.6702 & 0.7218 \\
 \hline
 $\tilde{\Gamma}$ & \textbf{0.0278} & 0.0175 & 0.0221 & \textbf{0.0373} & 0.0311 & 0.0279 \\
 \hline
\end{tabular}
\end{center}

\end{document}